\documentclass[]{aastex63}

\newcommand{\fnum}{erg cm$^{-2}$ s$^{-1}$}
\newcommand{\flam}{erg cm$^{-2}$ s$^{-1}$ $\rm{\AA}^{-1}$}

\usepackage{romannum}
\usepackage{wrapfig}
\usepackage{fancyvrb}
\VerbatimFootnotes
\usepackage{enumitem}                   
\usepackage{bm}
\shorttitle{RADYN Flare Heating Models}
\shortauthors{Kowalski et al.}
\usepackage{amsmath}
\usepackage{lineno}

\begin{document}

\title{Time-dependent Stellar Flare Models of Deep Atmospheric Heating}

\correspondingauthor{Adam F Kowalski}
\email{adam.f.kowalski@colorado.edu}

\author[0000-0001-7458-1176]{Adam F. Kowalski}
\affiliation{National Solar Observatory, University of Colorado Boulder, 3665 Discovery Drive, Boulder, CO 80303, USA}
\affiliation{Department of Astrophysical and Planetary Sciences, University of Colorado, Boulder, 2000 Colorado Ave, CO 80305, USA}
\affiliation{Laboratory for Atmospheric and Space Physics, University of Colorado Boulder, 3665 Discovery Drive, Boulder, CO 80303, USA.}

\author[0000-0003-4227-6809]{Joel C. Allred}
\affiliation{NASA Goddard Space Flight Center, Solar Physics Laboratory, Code 671, Greenbelt, MD 20771, USA}

\author[0000-0001-9218-3139]{Mats Carlsson}
\affiliation{Institute of Theoretical Astrophysics, University
              of Oslo, P.O. Box 1029 Blindern, 0315 Oslo, Norway}
              \affiliation{Rosseland Centre for Solar Physics, University of Oslo, P.O. Box 1029 Blindern, 0315 Oslo, Norway}

\begin{abstract}
Optical flares have been observed from magnetically active stars for many decades; unsurprisingly, the spectra and temporal evolution are  complicated.  For example, the shortcomings of optically thin, static slab models have long been recognized when confronted with the observations.  A less incorrect -- but equally simple -- phenomenological $T \approx 9000$ K blackbody model has instead been widely adopted in the absence of realistic (i.e., observationally-tested) time-dependent, atmospheric models that are readily available.   
We use the RADYN code to calculate a grid of 1D radiative-hydrodynamic stellar flare models that are driven by short pulses of electron-beam heating.  The flare heating rates in the low atmosphere vary over many orders of magnitude in the grid, and we show that the models with high-energy electron beams compare well to the global trends in flux ratios from impulsive-phase stellar flare, optical spectra. The models also match detailed spectral line shape properties.  We find that the pressure broadening and optical depths account for the broad components of the hydrogen Balmer $\gamma$ lines in a powerful flare with echelle spectra.  The self-consistent formation of the wings and nearby continuum level provide insight into how high-energy electron beam heating evolves from the impulsive to the gradual decay phase in white-light stellar flares.
The grid is  publicly available, and we discuss possible applications.
\end{abstract}


\section{Introduction} \label{sec:intro} 

Stellar flares are powerful bursts of radiation that are observed from radio frequencies through the X-rays.  They are thought to be caused by similar processes as in solar flares:  a catastrophic release of magnetic energy into the corona accelerates particles, which either escape to space or return the energy to the atmosphere as they thermalize.  The response of the stellar atmosphere initially occurs at lower altitudes within the chromosphere, which cools by radiating away this energy.  Early in a flare,  radiation can predominantly occur through optical and near-ultraviolet continua \citep{HP91, Woods2004, Hawley1995, Osten2015, Schmitt2019}, thus giving rise to a broadband visible-wavelength response that is sometimes referred to as the white light.  

 White-light flares dominate the short-duration variability in light curves of magnetically active stars in \emph{Kepler}, K2, and TESS data.  For example, the M dwarf GJ 1243 flares at a detectable level in \emph{Kepler} for $\approx$ 30\% of the time \citep{Hawley2014}.  The short-wavelength extensions of the optical and near-infrared continuum in these flares are important for assessments of exoplanet atmospheric photochemistry \citep{Loyd2018, Tilley2019, Segura2018} and surface biology \citep{Howard2020, Abrevaya2020}.  However, there is still not a self-consistent physical theory, from magnetic energy release to atmospheric heating, that explains the  spectral properties of stellar flare white-light radiation.  
 
 A phenomenological $T \approx 8500 - 14,000$ K blackbody model is consistent with some stellar flare spectra in the impulsive phase \citep{HF92, Fuhrmeister2008, Kowalski2013, Lalitha2013, Kowalski2016}, but other flare spectra show clear signatures of Balmer jumps and cooler blackbody temperatures in the optical \citep[e.g.,][]{Kowalski2019HST}. The optical color temperature attains the hottest values in the rise and peak phases \citep{Mochnacki1980} when the Balmer jumps are smallest \citep{Kowalski2013}. In the fast decay phase, the optical color temperatures rapidly decrease and settle to a cooler value around $5500$ K through the gradual decay phase \citep[see][for reviews]{Kowalski2018, Kowalski2024LRSP}.  The evolution of blackbody temperatures in stellar flares has been investigated with broadband photometry \citep{Hawley2003, Robinson2005, Zhilyaev2007, Howard2020}, but the interpretation of the continuum radiation can be difficult without spectra \citep{Allred2006, Kowalski2019HST}.  

 Large heating rates in gas-dynamic or radiative-hydrodynamic (RHD) models produce moderate-to-large optical depths at continuum wavelengths and have been successful in explaining some of the observed continuum properties \citep{Livshits1981, Katsova1981, Katsova1991, Katsova1997, Kowalski2015, KA18}.  However, the beam current densities may be too large to last during the propagation from the corona to the chromosphere.  Further, electron-beam-generated chromospheric condensations \citep[e.g.,][]{Livshits1981, Fisher1985VII, Fisher1989, Gan1992} in some of these models produce hydrogen Balmer line spectra that are too broad \citep{Kowalski2022Frontiers}; other recent perspectives and arguments have been provided by \citet{Belova2019} and \citet{Morchenko2020A, Morchenko2020B}. Stellar flare RHD models have considered smaller electron beam heating fluxes, but the model Balmer jumps are larger than in all known spectral observations; the photosphere is radiatively backheated by this Balmer continuum radiation, which originates at chromospheric heights \citep{Allred2006}.  Other models of stellar chromospheric flares include static NLTE slab calculations \citep{Jev1998, Garcia2002, Morchenko2015},  static  LTE slab calculations \citep{Kunkel1970,Eason1992,Heinzel2018,Simoes2024}, semi-empirical atmospheric adjustments with NLTE radiative transfer \citep{Houdebine1991, Christian2003, Fuhrmeister2005, Fuhrmeister2010, Kowalski2012SoPh, Schmidt2012}, and static energy equilibrium loop models with NLTE radiative transfer \citep{HF92}.  The semi-empirical model atmospheres of \cite{Cram1982} explored six variations of temperature and electron density adjustments within the deep chromosphere and photosphere, one of which produced very broad Balmer line wings and a hot optical blackbody color temperature.  For a detailed review of stellar flares and stellar flare modeling, see \citet{Kowalski2024LRSP}.

In this study, we present a grid of radiative-hydrodynamic stellar flare models that reproduce the range of observed optical and near-ultraviolet continuum and hydrogen Balmer emission line properties.  The atmospheres are driven by heating rates from power-law distributions of electrons.  On the Sun, nonthermal electrons can produce hard X-ray sources that are co-temporal and co-spatial with the white-light radiation at the footpoints of newly-reconnected magnetic loops \citep{Hudson1992,Neidig1993, Neidig1994,Xu2006, Fletcher2007,Kawate2016}.  However, the parameters of the power-law distributions in our grid are varied through a larger range than considered for solar flare models \citep[e.g.,][]{Carlsson2023} so that deep heating achieves optical blackbody color temperatures of $T > 10^4$ K in the emergent radiative flux spectra. The grid thus essentially follows on the static, semi-empirical model predictions of \citet{Cram1982} by including the evolution of self-consistent atmospheric opacities, flows, Coulomb heating rates, and non-equilibrium radiative transfer.  The detailed spectra and the hydrodynamic output are made publicly available to facilitate comparisons to and interpretations of stellar flare data.  The grid spectra have already been used for a wide range of purposes \citep{Kowalski2017Broadening, Namekata2020, Kowalski2022Frontiers, Kowalski2023, Brasseur2023, Howard2023, Monson2024}. 

The paper is organized as follows. Section \ref{sec:rhdmodels} contains an extensive description of the model setup, the grid of flare heating parameters, and the context of 1D flare loop models.
The analysis of the grid models (Section \ref{sec:analysis}) is divided into three parts.  Section \ref{sec:quantities} presents an overview of the time-averaged optical and NUV spectral quantities calculated from each model.  In Section \ref{sec:global}, the quantities from a subset of the grid are compared to global trends among impulsive-phase observations of dMe flares.  Section \ref{sec:detailed} demonstrates detailed modeling of the hydrogen Balmer $\gamma$ line in a remarkable dMe event whose broad lines have not yet been discussed in detail elsewhere.  We discuss the applicability and limitations of the model grid in Section \ref{sec:discussion}, and Section \ref{sec:conclusions} includes some concluding remarks.  The Appendix describes how to access the model output and additional documentation.

\section{Radiative-Hydrodynamic Flare Models} \label{sec:rhdmodels}

\subsection{Model Setup} \label{sec:basicsetup}
We calculate a grid of RHD models with the \texttt{RADYN} code \citep{Carlsson1992B, Carlsson1995, Carlsson1997, Carlsson2002, Allred2015, Carlsson2023}.  To simulate flare heating, we model the energy deposition from a power-law distribution (hereafter, ``beam'') of electrons.
The one-dimensional equations of mass, momentum, internal energy,  and charge conservation are solved on an adaptive grid \citep{Dorfi1987} with the equations of plane-parallel radiative transfer and level populations for a six-level hydrogen atom, a nine-level He I/II/III atom, and a six-level Ca II/III ion \citep[see][for details]{Allred2015, Carlsson2023}.  
The numerical method of solution consists of a semi-implicit (multi-dimensional Newton-Raphson) linearization and iteration of the time-centered spatial derivatives with an adjustable time step \citep{Dorfi1995, Carlsson1998, Dorfi1998, Abbett1998}.  An advantage of this method is that the time step is not limited to the CFL condition, which is very short in the flare transition region.  Time steps in the flare simulations in our grid often exceed the Courant step by factors of $\mathcal{O}(10^3)$, as expected \citep{Dorfi1998}, but they can also be very small and unwieldy.  

The model loop geometry is a quarter circle with an arc length of $l = 10$ Mm and a uniform cross-sectional area.  The surface gravity is log$_{10}$\ $g_{\star} /[$cm s$^{-2}]=4.75$, and the calculated effective temperature is $T_{\rm{eff}} \approx 3600$ K.   The loop apex has an electron density of $n_e \approx 3\times10^{10}$ cm$^{-3}$ and plasma temperature of $T \approx 5$ MK at $t=0$~s, which are in line with those determined from quiescent X-ray spectra of dMe stars \citep{Osten2006, Raassen2007, Liefke2008}.  Non-radiative heating is applied to the corona\footnote{We apply a constant energy deposition of $0.23$ erg cm$^{-3}$ s$^{-1}$ at $z \ge 460$ km, which causes thermal conduction to transition from a net cooling term at $z > 1.20$ Mm to a net heating term at $0.41 < z < 1.20$ Mm.} and to the photosphere to emulate convective energy flux \citep[e.g.,][]{Gustafsson2008}, which prevents catastrophic cooling before  flare heating begins.  The atmosphere begins in a self-consistent state of hydrostatic equilibrium, which is achieved through the steps that are described in the documentation associated with \citet{Carlsson2023}.   The ``height'' variable, $z$, is the arc distance along the loop, and  $z=0$ km corresponds to $\tau(\lambda=500\ \mathrm{nm}) = \tau_{500} = 1$ in the preflare state, where $\tau$ is the optical depth.
The one-dimensional formulation of the equations of hydrodynamics is justified if $\beta \ll 1$, where $\beta$ is the ratio of gas to magnetic pressure.
Our models are thus meant to represent the centers of very strong flux tubes in plage-like environments, where flare ribbons are often observed on the Sun.  
The model loop is relatively small and represents a compact, low-lying flare loop that appears early within a growing arcade of bright loops in EUV filtergrams \citep[e.g.,][]{Veronig2006, Liu2013, Chen2020}.  A reflecting upper boundary emulates waves originating from an identical conjugate footpoint of the loop.  Thus, the flare loops are symmetric with respect to the apex.  We also impose a $v_z = 0$ lower boundary condition \citep[as in the F-CHROMA solar flare models;][]{Carlsson2023}, which does not affect the chromospheric dynamics over the short timescales ($\Delta t \le 10$~s) 
covered by these simulations.  An interesting effect due to the reflected waves at the upper boundary is discussed in Section \ref{sec:quantities}.

  An electron beam is injected at the loop apex to simulate flare heating.  The electron beam pitch-angles are forward-Gaussian distributed with respect to the loop axis.  The volumetric flare heating rate is determined by the steady-state solution of the distribution function of energy, pitch angle, and position along the loop.  The diffusion and drag of beam particles due to Coulomb collisions with the ionized and neutral components of the background (thermal) plasma are included through the Fokker-Planck formulation, which is solved in a module that was further developed into the standalone \texttt{FP} code that solves the distribution function with the return current force for any beam particle mass and charge \citep{Allred2020}. The heating rate is the gradient of the beam energy flux  \citep[see][]{Allred2015}.  The heating rate is calculated at every time step in the radiative-hydrodynamic simulation because gradients of the atmospheric density  evolve rapidly after $t=0$~s in response to the flare heating.  Collisional excitation and ionization rates of hydrogen \citep{Fang1993} and helium \citep{AR85} with the electron beam are included in the detailed rate equation following \citet{Allred2015}, but these are rapidly overwhelmed by the thermal collision rates in our simulations.  In this first generation of models, the collective force term \citep[i.e., the third term in Eq.\ 1 of][]{Allred2020} that accounts for return current electric fields and magnetic mirroring has not yet been included.  These physical processes warrant detailed study that is outside the scope of this work, but some of these simplifications are broadly discussed in Section \ref{sec:discussion}.

  M dwarf flares produce copious radio and millimeter emission, which is usually constrained to be gyrosynchrotron or synchrotron radiation from a population of mildly relativistic electrons \citep{Gudel1996, Smith2005, Osten2005, MacGregor2018}.
 All of the models in the grid are calculated for injected number-flux distributions with small (``hard'') power-law indices in the range of $\delta = 2.5 - 4$, which is consistent with the interpretation of available stellar flare constraints \citep[e.g.,][]{Smith2005, Osten2007, MacGregor2020, MacGregor2021}.  The peak injected beam energy flux densities (hereafter, we refer to the flux density as the ``flux'') above a low-energy cutoff, $E_c$, span  four orders of magnitude: $10^{10}$ (F10), $10^{11}$ (F11), $10^{12}$ (F12), and $10^{13}$ (F13) erg cm$^{-2}$ s$^{-1}$.   Our grid of M dwarf flare models includes a large range of low-energy cutoff values:  $E_c = 17, 25, 37, 85, 150, 200, 350, $ and $500$ keV.   The large beam flux densities and high low-energy cutoffs in the grid are necessary to produce deep chromospheric heating in M-dwarf flares.  The energy fluxes of $10^{12}-10^{13}$ erg cm$^{-2}$ s$^{-1}$ are consistent with the largest values above $E_c \approx 20$ keV that have been inferred through the collisional thick-target model (CTTM) interpretation of some solar flare hard X-ray events \citep{McClymont1986, Canfield1991, Wulser1992, Neidig1993, White2003, Krucker2011}. Large, low-energy cutoffs, $E_c \approx 70-120$ keV, are also occasionally inferred in CTTM modeling \citep{Holman2003, Warmuth2009}.   Furthermore, recent numerical advances \citep{Arnold2021} generate $\delta =3$ electron-beam power-law distributions that contain a large percentage, up to $\approx 20- 30$\%, of the ambient coronal density.  Note that  coronal electron densities in M dwarf atmospheres outside of flares \citep[e.g.,][]{Osten2006, Liefke2008}  are even larger than fiducial solar coronal values ($n_e \approx 10^{9}$ cm$^{-3}$). Section \ref{sec:injection} further discusses  the energetic requirements of high-energy electron beams in stellar flares.

The time-evolution of the electron beam energy injection into a loop is poorly constrained in M dwarf flares, due in part to the lack of direct spatial resolution. However, the injection of high-levels of nonthermal energy into any given flare loop is probably very short in order to be consistent with shortest rise times of $\approx 2 - 10$~s in optical, $U$-band, and UV observations \citep{Moffett1974, Robinson1995, Mathioudakis2006, Kowalski2016,  MacGregor2021}. We choose $\Delta t = 2.3$~s as a representative timescale for injection into the \texttt{RADYN} model loop; this is sufficiently long compared to time-of-flight differences between the high- and low-energy electrons in the beams. The limitations of the applicability of these models due to the choice of short beam injection timescales are discussed further in Section \ref{sec:discussion}.  The grid consists of two types of time profiles for the injected beam flux:  a constant injection rate and a ramp up/down injection rate.  The constant injection models are helpful for isolating effects due solely to the atmospheric evolution, and the ramp up/down evolution is physically motivated by the pulsed-injection model of solar flare hard X-ray bursts \citep{Aschwanden2004}. We now discuss our application of pulsed-beam injection to stellar flares.  

\subsection{Pulsed Beam Injection} \label{sec:injection}
There are a large number of assumptions that are needed to simulate the RHD of a stellar flare.  Although neither the magnetic field dynamics nor the particle acceleration process are included in our RHD simulations,  we relate some of the assumptions in our grid to a larger geometrical context.  This provides some insight into the applicability and limitations of 1D RHD flare loop modeling.  We draw on state-of-the-art advancements from models of magnetic field geometries in solar flares in order to further motivate the extreme ends of  electron beam properties within the grid.

There are many observed timescales of nonthermal hard X-ray variations in solar flares, from tens of milliseconds to tens of seconds \citep{Kiplinger1984}.  However, it is not yet clear how each of the respective beam populations contribute to the total thermal response of the chromosphere.  Indeed, phenomenological power-law, power-spectra of frequencies adequately explain much of the temporal variations in solar and stellar flares \citep{Inglis2015}.  The slow and fast components of solar flare light curves have also been attributed to a ``trap-plus-precipitation'' model of hard X-rays \citep{Melrose1976, Aschwanden1998}. 
At high temporal resolution, $\Delta t \approx 4-20$~s bursts  \citep{Aschwanden1998, Xu2006, Penn2016, Rubio2016, Kawate2016, Collier2023} and a long-duration, smoothly varying component are often present in hard X-ray and optical solar flares.  Even shorter,  $\Delta t \approx 0.1-3$~s, bursts are observed in H$\alpha$ and hard X-ray solar flare kernels \citep{Kiplinger1984, Aschwanden1995, Aschwanden1998, Aschwanden1998Wavelet, Aschwanden2004, Qiu2012, Radz2007, Radz2011}.  A large fraction of the hard X-ray emission is often contained in the smoothly varying component \citep{Aschwanden1998B}, which may originate from a population of electrons that are trapped in  loops due to magnetic field convergence at the chromospheric footpoints. Trapping may also occur through some type of plasma-confining effect higher up in the coronal loop volume \citep[e.g.,][]{Li2013, Egedal2015, Fleishman2022}. In some solar flares, however, there is little evidence for large populations of trapped particles, and both the hard X-rays and gyrosynchrotron radiation can thus exhibit similarly pulsed temporal variations \citep[][see also \cite{Kundu2001} and \citealt{Krucker2020}]{White2003}.  A wide variety of pulsed injection schemes have been used to drive models of solar flares \citep{Emslie1983, Lee2000, Kasparova2007, Doyle2012, Rubio2016, Simoes2017, Carlsson2023}.

 A step function is the simplest type of pulsed injection.  However, we also desire a scheme that exhibits a gradual decrease in the flux injected at the top of a model \texttt{RADYN} loop.
The dynamic pulsed injection model of \citet{Aschwanden2004} assumes adiabatic\footnote{Meaning that the pitch angles change as the particles  bounce around in this region, while the total energy of each particle remains constant according to the adiabatic invariants.  The so-called betatron heating and acceleration \citep[e.g.,][]{Veronig2006, Birn2017} are not considered.} particle motion as the magnetic field lines retract after their reconnection.  It is a parametrized model that accounts for the increase in magnetic field from the reconnection X-point to the  previously-reconnected looptops that are downstream of the reconnection outflow.   
As the ratio of the magnetic field in the retracting loop to that in the underlying looptops decreases, the angular size of the loss cone increases, which allows particles with larger pitch angles to escape out of the trap.  These particles precipitate into the lower regions of the loops (footpoints) and deposit the energy that drives the flare in the chromosphere. Footpoint convergence of the magnetic field causes further trapping of electrons before they reach the chromosphere, but this effect is not considered in this grid of models.  We assume parameter values for the pulse full-width-at-half maximum (FWHM), $t_w = t_{\rm{FWHM}} = 2.3$~s, the height of the fully contracted looptops, $h_L=0.64 \times 10^9$ cm, and the initial height of the contracting fields, $h_Y$ (which is $h_X$ in \citealt{Aschwanden2004}), where we assume that $h_Y = 2 h_L$.  The parameter $q_a$ is equal to $t_w$, which is the width of the flux injected at the \texttt{RADYN} looptop, divided by a timescale for the injection of particles from the acceleration region. For the models here, we assume that $q_a=1.28$, which gives a cumulative number of electrons injected into a retracting  loop from the acceleration region that is linearly proportional to $t$.

One can further obtain an estimate of the low-energy cutoff,  $E_c$, to the nonthermal distribution from recent numerical simulations \citep{Haggerty2015} according to 

\begin{equation} \label{eq:haggerty}
  E_{\rm{c}} \approx \frac{0.15}{2} m_i c_{A, X}^2  \approx 0.15 \frac{B_X^2}{8 \pi n_e},
\end{equation}

\noindent where $B_X$ is the magnetic field at the reconnection X-point.  For $B_X = 1$ kG and using the \texttt{RADYN} model loop apex electron density, Eq.\ \ref{eq:haggerty} gives $E_c \approx 125$ keV.   Here, it is assumed that the retraction occurs at the Alfven speed, $c_{A,X}$, upstream of the reconnection outflow.  Detailed calculations show deviations from an Alfvenic speed retraction process \citep[e.g.][]{Longcope2018}, and energy transfer with various turbulent and shock phenomena may also occur farther downstream \citep{Somov1997, Li2012, Kontar2012, Egedal2015, Kontar2017, Ruan2023}.  Aside from our chosen value of $t_w = 2.3$~s (see Section \ref{sec:discussion} for further discussion of the chosen timescale and loop sizescales), the values of the parameters from the pulsed injection model are not explored here.
There are several ways one can also estimate the energy flux that powers a flare loop.  We use a formula for the energy that is released through field line shortening \citep{Longcope2016}, which is 

\begin{equation}
F_{fl} \approx \frac{1}{2\pi} B_X^2 c_{A,X} \sin^4(\Delta \theta / 4) \approx 0.9 - 5\times 10^{13}\ \mathrm{erg}\ \mathrm{cm}^{-2}\ \mathrm{s}^{-1}
\end{equation}

\noindent This range of flare fluxes corresponds to a range of shear angles, $\Delta \theta$, which are proportional to the magnetic field components that are parallel to the active region polarity inversion line. These first-principle assessments, which draw from state-of-the-art numerical modeling, together demonstrate that high beam fluxes and large low-energy cutoffs are not within the realm of fantasy in stellar flares that occur in a region with $\approx$ kG fields in the low corona.

\vspace{3mm}

\begin{figure}
\begin{center}
\includegraphics[scale=.33]{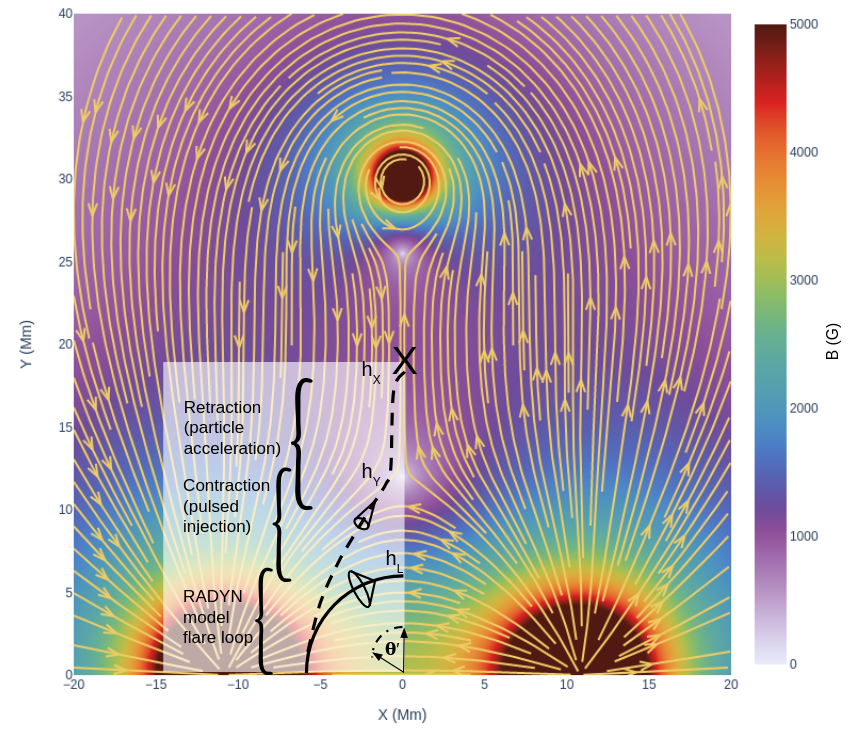}
\caption{ Magnetic field vector lines from the 2D analytic loss-of-equilibrium model described in \citet{Reeves2005}.  
The reconnection point (``X'') is assumed to occur halfway up the current sheet where the magnetic field strength is $\approx 1$ kG in this particular configuration.  Three regions to the analogous diagram in \citet{Chen2020} are indicated.  The loss cone opening angle increases as the fields contract toward the flare arcade below, allowing electrons with pitch angles within the cone limits to precipitate from the contracting region. The \texttt{RADYN} modeling simplifies the geometry by assuming a quarter-circle loop of constant area and a gravitational acceleration along the loop axis that varies according to $g_z = g_{\star} \sin(\theta^{\prime})$.
  In reality, the three  regions are not as distinct as drawn here, and the dynamics cannot be shown accurately with a single snapshot  \citep[see][for the evolution of fields in a similar geometrical configuration]{Chen2020}.  The streamlines show the magnetic field direction, while the background color scale shows the magnetic flux density. 
\label{fig:geometry}}
\end{center}
\end{figure}

Figure \ref{fig:geometry} is an illustration of a possible magnetic field geometry in our application of the pulsed injection model.  The reconnecting magnetic field geometry is similar to the analytic \citep[][see also \citealt{Reeves2007}]{Forbes1995, Lin2000, Reeves2005} and numerical \citep{Athena2008, Athena2020} models analyzed in \citet{Chen2020}.  Large magnetic fields are assumed in the reconnecting current sheet region, but the field strengths are not that much larger in the figure than those inferred in an off-limb solar flare \citep{Chen2020}.  The implied photospheric fields are also very strong, but large fields of $B \approx 8-12$ kG, may cover $\approx 10-20$\% of the surfaces of magnetically active M dwarfs \citep{Shulyak2019, Reiners2022}.  
Within this geometry, \cite{Chen2020} indicate regions that correspond to the ``reconnecting current sheet'', ``contracting loops'', and ``flare arcade''\footnote{In a different off-limb flare, \citealt{Liu2013} discovered ``fast downward contractions'' and  ``slow downward shrinkages'', which are most easily seen after the hard X-ray impulsive phase.}   In our envisioned scenario,  the pulsed injection originates in the contracting loop regime rather than in the retracting loops within the vicinity of the reconnecting current sheet region \citep[where the bulk of particle acceleration plausibly occurs;][]{Arnold2021}.   One should keep in mind that there is actually very little agreement on the details in the dynamic processes and energy flow within this type of scenario \citep[e.g.,][]{Fletcher2008}, but many of the important features are consistent with state-of-the-art observations and 2.5/3D MHD modeling of eruptive events \citep{Sui2003, Sun2012, Liu2013, Longcope2018, Krucker2020, Chen2020, Rempel2023, Volpara2024}.  The extensions to stellar coronae are even less established, including whether such models of mass ejections are appropriate \citep{Crosley2018A, Alvarado2018}, possibly due to strong overarching fields as in confined solar flares \citep{Thalmann2015}.  Nonetheless, Figure \ref{fig:geometry} is an adequate starting framework for linking some properties of the \texttt{RADYN} model spectra to particle acceleration and the action of strong, low-lying coronal magnetic fields that are expected in young M-dwarf flare stars.

\begin{figure}
\begin{center}
\includegraphics[scale=1.0]{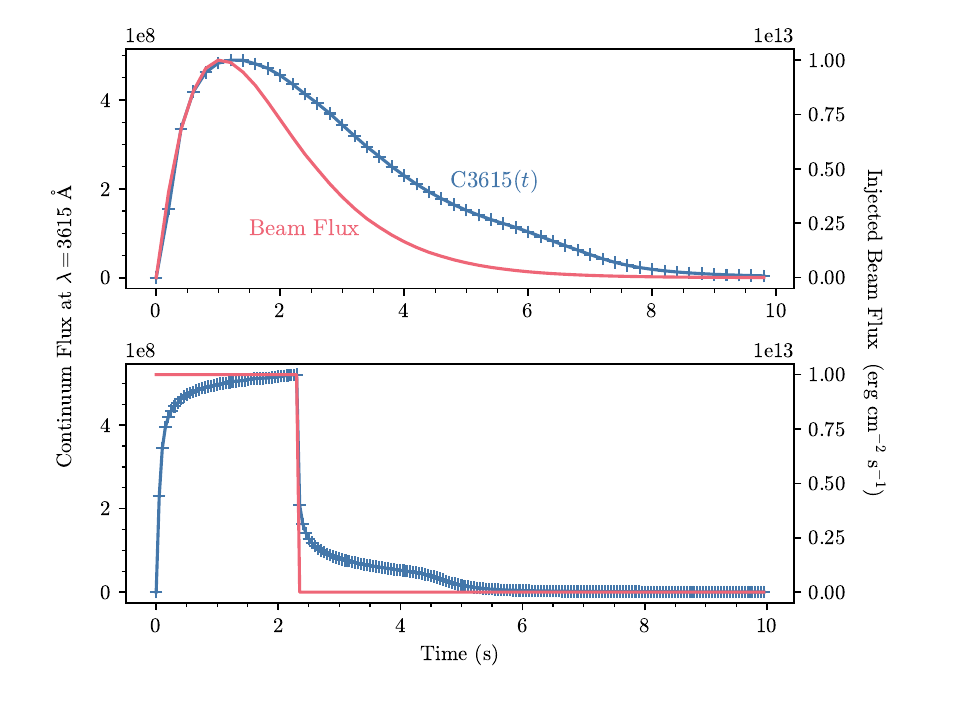}
\caption{ Evolution of electron beam flux injected at the \texttt{RADYN} model loop top for two types of pulses in the model grid: (\emph{Top}) pulsed ramp up/down precipitation flux from the formulae in \citet{Aschwanden2004} in the \texttt{mF13-85-3} model at $\Delta t = 0.2$~s; (\emph{Bottom}) a pulsed step-function injection in the \texttt{cF13-85-3} model at  $\Delta t = 0.05$~s.  The evolution of the emergent radiative flux at a Balmer continuum wavelength is shown in each panel for comparison. 
\label{fig:Figure_BeamEvol}}
\end{center}
\end{figure}

An example of pulsed, ramp up/down beam energy injection into a \texttt{RADYN} model loop is shown in Figure \ref{fig:Figure_BeamEvol} using the parameter $t_{\rm{FWHM}}=2.3$~s of the \citet{Aschwanden2004} model as described above.  This pulse is normalized to a maximum injected flux density of $10^{13}$ \fnum , the low-energy cutoff is $E_c=85$ keV, and the power-law index is $\delta=3$.  We introduce the nomenclature for unique model identification as \texttt{mF13-85-3}, where \texttt{m} means that the electron beam flux that is indicated by the $F$-number is a maximum over the pulsed, ramp up/down injection.  Note that time-integrated energy in the pulse in the top panel is a factor of 2.5 larger than the peak flux.   Alternatively, \texttt{c} indicates a constant injected flux pulse at $t=0-2.3$~s and zero flux at $t > 2.3$~s (unless otherwise indicated in the model identifier, the injection timescale is $\Delta t = 2.3$~s).  In Figure \ref{fig:Figure_BeamEvol}, we also show the evolution of the detailed emergent radiative flux at a representative wavelength, $\lambda = 3615$ \AA, that is just shortward of the ideal Balmer recombination limit ($\lambda = 3646$ \AA).  The evolution of the continuum flux indicates that the thermal radiation from the chromosphere responds over slightly longer timescales than the injected nonthermal energy \citep{AllredPhD, Simoes2017}.

\subsection{Emergent Radiative Flux Spectra} \label{sec:spectracalc}
\texttt{RADYN} provides spectra from $\lambda = 6-40,000$ \AA\ in physical (cgs) units, which facilitates comparisons to flux-calibrated stellar observations.
Emergent radiative surface flux density spectra (\flam; hereafter, just ``flux'') are calculated within \texttt{RADYN} using a Gaussian quadrature sum of the emergent intensity at five $\mu$ values ($\mu = 0.05, 0.23, 0.50, 0.77, 0.95$, where $\mu = \cos (\theta)$ and $\theta$ is the angle with respect to the atmospheric normal), and at 95 continuum wavelengths.  We denote the flux at a specific continuum wavelength $\lambda$ as ``C$\lambda (t)$''.  Throughout, a prime symbol indicates that the preflare flux is subtracted from the flux at time $t$.

\begin{figure}
\includegraphics[scale=1.0]{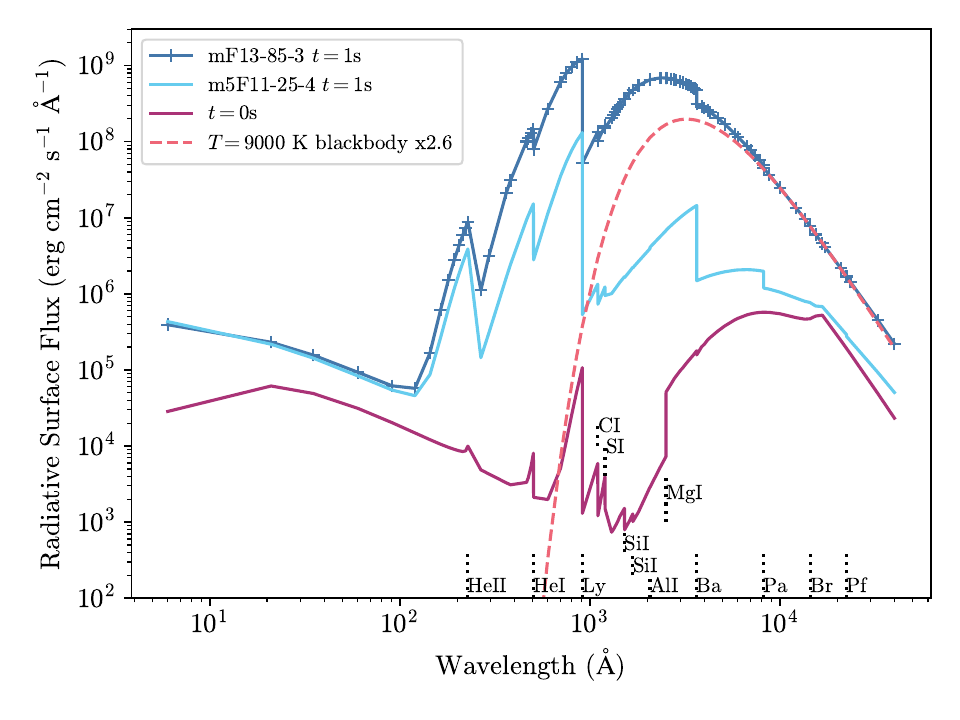}
\caption{ Radiative surface flux spectra from a representative range of heating rates covered by the flare models in the grid.  The pre-flare surface flux spectrum from \texttt{RADYN} and a $T=9000$ K blackbody function (scaled by a factor of 2.6) are shown for comparison.   The bound-free continuum opacities that generate the ultraviolet edges \citep{Vernazza1976} in these spectra are due to the following:  C I $\lambda < 1100$ \AA\ ($2p^2\ ^3P$), S I $\lambda < 1199$ \AA, Si I $\lambda < 1525$ \AA\ ($3p^2\ ^3P$), Si I $\lambda< 1682$ \AA\ ($3p^2\ ^1D$), Al I $\lambda < 2076$ \AA\ ($3p^2\ P^0$), and Mg I $\lambda < 2513$ \AA\ ($3p\ ^3P^0$).  The detailed bound-free edges are due to He II $\lambda < 227$ \AA\ and He I $\lambda < 504$ \AA\ in addition to the typical H I edges.
\label{fig:ContSpec}}
\end{figure}

A representative range of detailed continuum spectra from the grid is shown in Figure \ref{fig:ContSpec} for two models, \texttt{mF13-85-3} and \texttt{m5F11-25-4}.  Bound-free discontinuities are present in the spectra because dissolved-level bound-free opacities \citep{Dappen1987, Hubeny1994, Hubeny2014} are not included in \texttt{RADYN} calculations.   Away from the bound-free edges, the optical color temperatures, the Balmer jump strengths, and the Balmer continuum slopes vary markedly between the \texttt{mF13-85-3} and \texttt{m5F11-25-4} models, which respectively heat column masses of  $\log_{10} m_c$ /[g cm$^{-2}$] $\approx -2$ and $-3$ to gas temperatures of $T\gtrsim 10^4$ K. It is not  unexpected that the major discriminatory power between these models occurs in the NUV and optical, while the far-ultraviolet (FUV) and X-ray and extreme-ultraviolet (XEUV) spectral regions are similar in shapes.  The color temperatures that are inferred from the X-rays at  $\lambda < 60$ \AA, for example, are comparable between these two models (Section \ref{sec:quantities}).  Figure \ref{fig:ContSpec} also shows how these models improve upon phenomenological, $T = 9000$ K blackbody modeling of white-light stellar flares.  

Background continuum opacity from elements other than those treated in detail (Section \ref{sec:basicsetup}) have been included assuming LTE, with photoionization cross-section data compiled from The Opacity Project dataBASE (TOPBASE\footnote{http://cdsarc.u-strasbg.fr}). 
The background opacities produce many of the faint bound-free edges in the spectra of Figure \ref{fig:ContSpec}.  The H$^{-}$ ion is not included in detail, but its population densities are self-consistent with the non-equilibrium electron densities in order to give accurate bound-free and free-free opacities.   The Ca I populations are also larger than in the lower solar atmosphere.  However, only a model Ca II ion file (with a Ca III ionization stage) is available in the detailed calculations.  Thus, we consider charge conservation using electrons from the LTE ionization balance of Ca I, II, and III, as for the other background elements, instead of with hydrogen and helium.   The bound-free opacities of hydrogen from $n = 6 - 8$ are included assuming LTE populations relative to the continuum.

 In the calculations, the population densities of H I are reduced by including LTE chemical dissociation equilibrium of H$_2$ \citep{Vardya1965, Tsuji1973}.  In the pre-flare photosphere, the reduction of H I is about 50\%.  Although this affects the H opacity, the models do not have molecular opacities directly included in detail or in the background.  Of course, M-dwarfs with $T_{\rm{eff}} \approx 3600$ K show prominent molecular absorption bands of TiO, MgH, and CaH  in their non-flare optical spectra \citep[\emph{cf.} Fig 4 of][]{Bochanski2007}.  It is reasonable to assume that the heating rates in the low chromosphere in our flare models are too large, and the atmospheres heat up too quickly, for the energies of molecular dissociation to be a limiting term in the internal energy equation. 

Higher density wavelength grids are used for the calculations across the emission lines of hydrogen, He I, He II, and Ca II that are treated in detail.  The hydrogen Balmer line spectra (H$\alpha$, H$\beta$, H$\gamma$) are modeled using the Doppler-convolved, TB09+HM88 line profile functions \citep{Vidal1970, Vidal1971, Vidal1973, HM88, Tremblay2009}, which accurately capture the pressure broadening from ambient, thermal electrons and ions in the density regimes of flare chromospheres \citep{Kowalski2017Broadening, Kowalski2022}. The pressure broadening of other Paschen and Lyman hydrogen lines is included as described in \citet{Kowalski2022}.  Within \texttt{RADYN}, the Balmer H$\alpha$, H$\beta$, and H$\gamma$ lines are calculated on frequency grids with 51, 31, and 31 points, respectively.  The sampling in the far wings ($|\lambda - \lambda_0| \gtrsim 2$ \AA) is courser than around the rest wavelength.   To mitigate systematic errors in the line-integrated fluxes of the H$\gamma$ line, we follow \cite{Kowalski2022Frontiers} and use a standard Feautrier solver to recalculate the emergent surface flux spectra on a 327 point wavelength grid with the frequency-independent, non-LTE source function from \texttt{RADYN} \citep[see the discussion in][]{Kowalski2022}.  We use a bilinear interpolation of $\log_{10} \phi_{\alpha}$ on a grid of $T$ and $n_e$,  where $\alpha$ is the typical detuning parameter $\propto |\lambda - \lambda_0|$, and $\phi_{\alpha}$ is the normalized line profile function.  This is  followed by a  four-point, third-order interpolation scheme on a grid of $\log_{10} \alpha$ and $\log_{10} \phi_{\alpha}$ \citep{Vidal1973}.

 Several of the low-order Balmer, Paschen, and Brackett lines are available in the model output (see Appendix).   We  provide high- and low-resolution spectra of H$\gamma$ and low-resolution spectra of the Balmer H$\beta$ and H$\alpha$ lines.  The diagnostic advantages of Balmer H$\gamma$ have been discussed elsewhere \citep{Kowalski2022} (see also Section \ref{sec:global}).  Additionally, new line-shape calculations \citep{Cho2022} that use the \emph{Xenomorph} code  \citep{Gomez2016} have shown that (at high density) the static approximation to the perturbing ions in a plasma and the time-ordering of collisions are more important for the H$\alpha$ line than have been previously demonstrated \citep{Stehle1994}.  However, the new line shape calculations of H$\gamma$ are essentially identical to the TB09+HM88 profiles that we use.  Extending the \emph{Xenomorph} calculations of  H$\alpha$ to lower electron densities, which are relevant for flare atmospheres, will be the subject of future work (Kowalski \& Gomez, in prep).

\subsection{XUV Backheating}

X-ray and ultraviolet (XUV; $\lambda = 1-2500$ \AA) backwarming is included following earlier \texttt{RADYN} models \citep{Abbett1999, Allred2005, Kowalski2015} that use the methods of \citet{Gan1990}, \citet{HF92}, and \citet{HF94}.  An incident radiation field is calculated from the Astrophysical Plasma Emission Code (APEC), which is part of AtomDB \citep{Smith2001}.  Photoionization rates from CHIANTI are used for the calculations of volumetric heating rates.  The capability to include the incident radiation field in the equation of radiative transfer \citep{Allred2015} was not yet incorporated into the M-dwarf version of \texttt{RADYN} when this grid was created.  However, the results of \citet{Allred2006} lead us to believe that differences among the treatments of XUV backheating are negligible in comparison to the large electron beam heating rates in this grid\footnote{Notably, however, the preflare emergent intensities of the Balmer lines are factors of $2-3$ fainter when the starting atmosphere is relaxed with the method of \cite{Allred2015}.}.

\subsection{Optically Thin Radiative Loss Function}

\begin{figure*}
\gridline{\fig{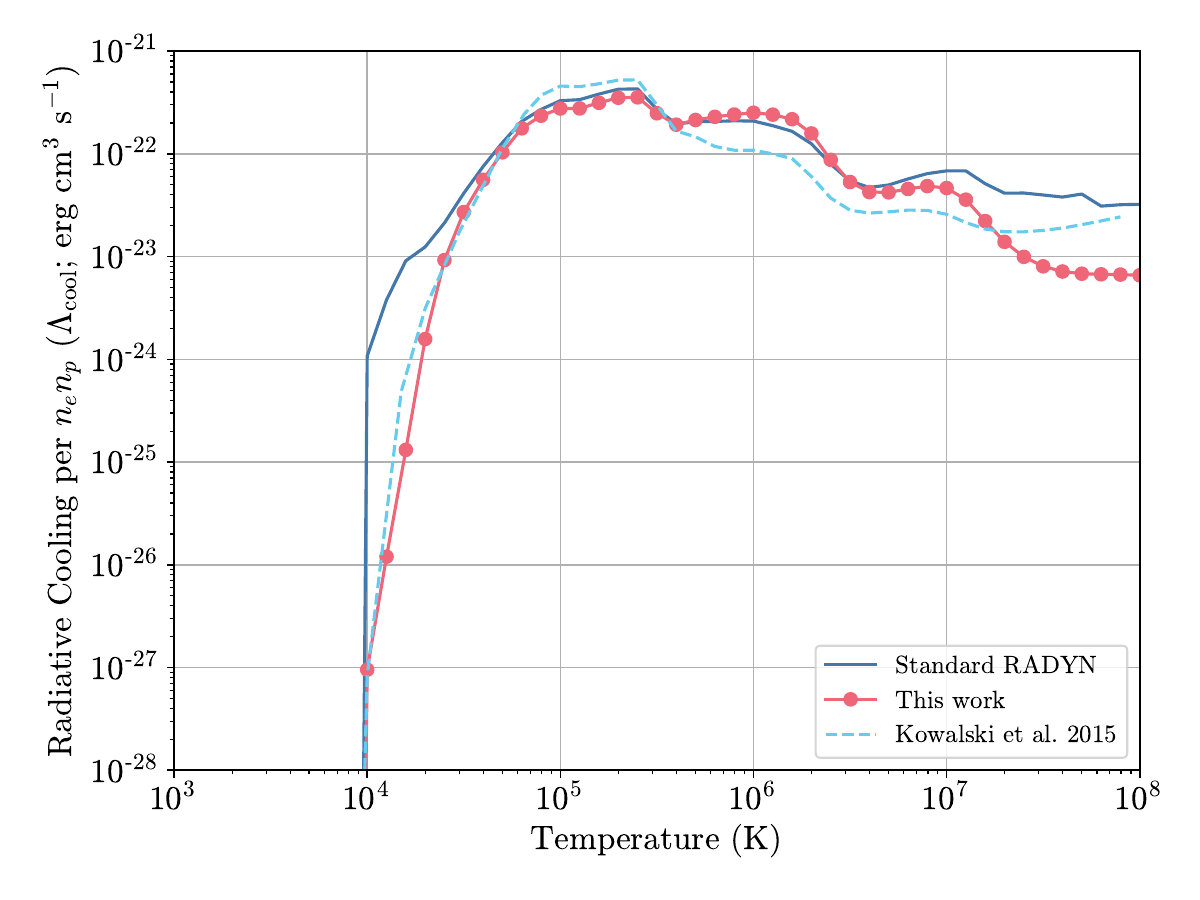}{0.45\textwidth}{(a)}
          \fig{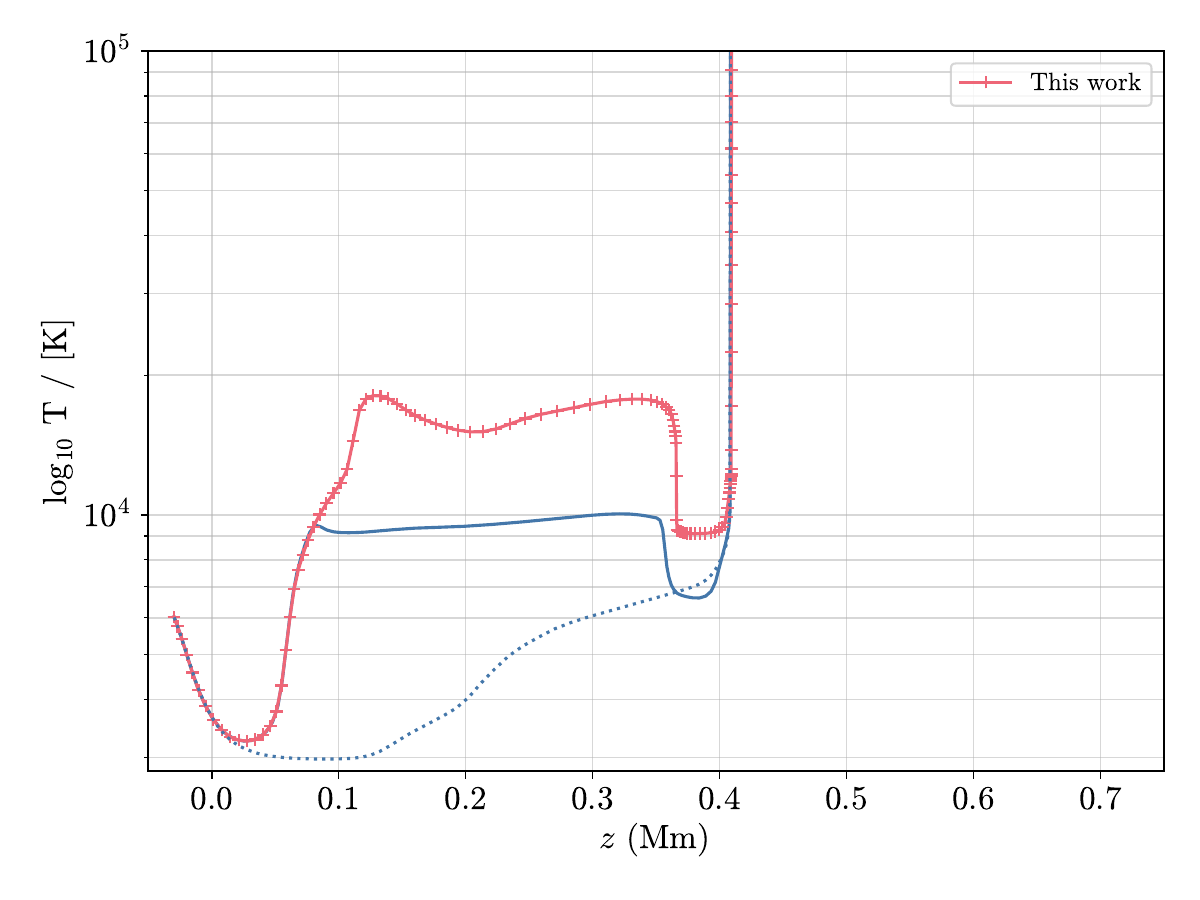}{0.45\textwidth}{(b)}
          }
\gridline{\fig{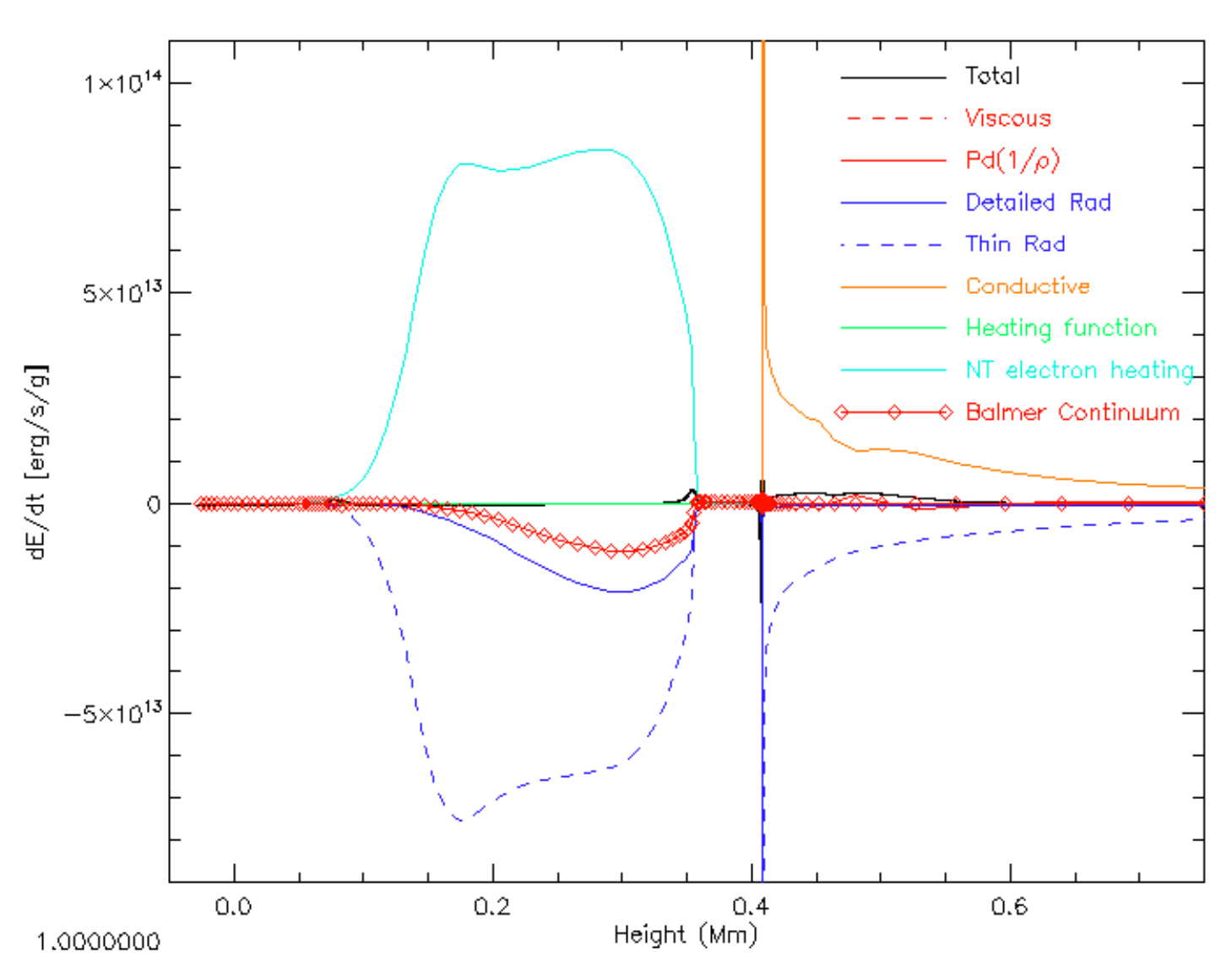}{0.45\textwidth}{(c)}
          \fig{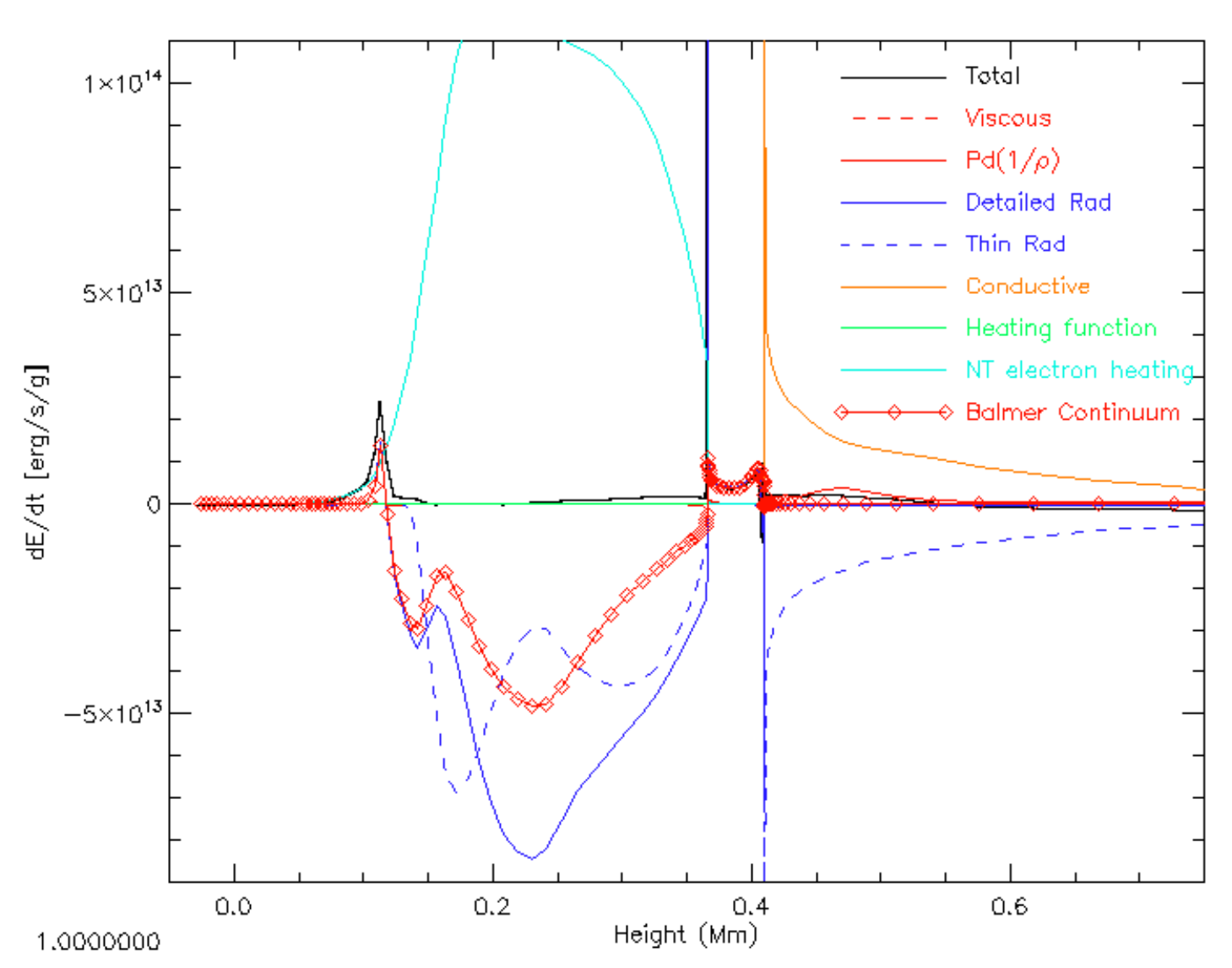}{0.45\textwidth}{(d)}
          }
\caption{ (a) Comparison of the optically thin loss function that is used in this work, the one used in \citet{Kowalski2015}, and one of the standard versions from the \texttt{RADYN} code.  (b)  Calculations of the \texttt{mF13-500-3} model gas temperatures using the standard thin loss function compared to this work. The temperature snapshots are obtained at $t=1$~s and correspond to the height ranges in the bottom panels.  The preflare temperature is shown as the dotted line.  Panels (c) and (d) show net contributions to the changes of internal energy per gram.  Individual flux divergence terms (divided by mass density) in the internal energy equation are labeled. The thin loss function in the grid is used in the snapshot in (d), and a standard optically thin loss function is used in (c).  Note the large differences between the detailed radiative cooling (which is dominated by wavelengths covering the Balmer continuum range) and the thin radiative cooling in the low atmosphere.  The model snapshots in (c) and (d) are the same as in panel (b).
\label{fig:TRLcompare}}
\end{figure*}

The optically thin radiative loss function, $\Lambda_{\rm{cool}}$, is an important term in the conservation equation for the internal energy in flare modeling.  
We use an optically thin loss function from CHIANTI version 8.0.1 \citep{Dere1997, DelZanna2015} to account for radiative losses at $T = 10^4 - 10^8$ K from  species that are not treated in detail.   We also exclude species for which the optically thin approximation is likely inaccurate in stellar flare atmospheres at low temperatures, $T \approx 10^4$ K.  Table \ref{table:thinloss} lists all of the atoms and ions that are excluded from the adopted thin loss function, which is shown in Figure \ref{fig:TRLcompare}(a).  A standard cooling curve that includes these species predicts $\approx 10^3$ greater radiative losses at the lowest temperature.  A similar cooling curve at low temperatures was used in the \texttt{RADYN} flare simulations described in \cite{Kowalski2015}.

The differences at high temperature, $T \gtrsim 10^7$ K, in Figure \ref{fig:TRLcompare} are partially due to excluding the free-free cooling from H II, He II, and He III in the thin losses.   In our experience with the grid of models,  the thermal conduction, which is of the classical Spitzer form with a flux-limited saturation value \citep{Smith1980} in \texttt{RADYN},  is much more important than optically thin radiative losses in the energy equation within the model flare coronae. The treatment of free-free loss terms is even less important in the evolution of the lower, cooler atmospheric regions (the focus of our study) in flare models. We also note that the cooling curve is dependent on the assumed abundances (\emph{cf.} Section 7.19 of the CHIANTI manual), which are discussed below.  The differences among the loss functions at low temperature, $T \approx 10^4$ K, originate primarily\footnote{In decreasing order, the six largest contributions to the bound-free radiative cooling at $T=10^4$ K are H I, Si I, Mg I, S I, C I, and Al I.  In decreasing order, the six  largest contributions to the bound-bound radiative cooling at $T=10^4$ K are H I, Mg II, Fe II, Ca II, Si II, and Al II.  Note, the bound-bound cooling is generally much larger than the bound-free. The main contributions to the remaining radiative cooling in our adopted thin loss function at $T=10^4$ K are bound-bound transitions in N I, S II, and Ni II in decreasing order.  }  from including or excluding bound-bound transitions in Mg II, Fe II, and Si II ions, which we now further describe and justify.

The lower-atmospheric temperatures (Figure \ref{fig:TRLcompare}(b)) in flare simulations are dramatically different between two versions of the thin loss function with the same injected beam (\texttt{mF13-500-3}).  Figures \ref{fig:TRLcompare}(c)-(d) show the contributions to the internal energy equation in the calculations of panel (b).  The thin cooling prevents a temperature increase above $T \approx 10^4$ K at large column masses and low altitudes in the atmosphere.  The emergent continuum flux spectra (not shown) are also quite different.   Thus, our choice of excluding many transitions altogether (Table \ref{table:thinloss}) from the loss curve at $T = 10,000 - 20,000$ K warrants further justification.  We compare emergent LTE intensity spectra for a bright and a faint Fe II emission line at $t=1$~s in the \texttt{mF13-85-3} model.  One calculation for each line includes the optical depth from the bound-bound transition (``optically thick calculation''), while another calculation excludes the bound-bound contribution toward the total optical depth over the wavelengths of the line (``optically thin calculation'').  The bound-bound emissivity is considered in LTE in both calculations, as are bound-free emissivities and opacities.  For the bright Fe II $\lambda 2632$ 
emission line, the emergent wavelength-integrated emergent intensity is $\approx 200$x greater than the optically thick calculation.  For the faint Fe II $\lambda 2814.445$ line, the intensity is a factor of $\approx 10-15$ larger than in the optically thick calculation.  These differences are rather striking and suggest that excluding low-temperature emission lines altogether from the thin loss table, as we have done, is warranted until a more rigorous approach \citep[e.g., as calculated in][for quiet Sun conditions]{Carlsson2012} is possible.

Additional assumptions are that the abundances constantly remain at solar coronal levels \citep{Schmelz2012} and the emissivities in the loss function are calculated from ionization equilibrium (in CHIANTI).  These assumptions probably do not hold in flares \citep[e.g.,][]{Colgan2008, Doyle2012, Warren2014}. We argue that the lack of uniform patterns in abundance changes among M-dwarf flares \citep[e.g.,][]{Osten2005, Paudel2021}, combined with the difficulties associated with metallicity diagnostics of M dwarfs \citep[e.g.,][]{Mann2013}, does not warrant varying the abundances 
 away from the solar values at this time.  Non-equilibrium rates of the elements not treated in detail may be important, but they must also be considered in the context of large densities that are produced in our models, which may in turn result in non-negligible optical depths in transition region lines \citep[e.g.,][]{Mathioudakis1999, Kerr2019Si}.  It would also not be surprising if non-LTE photo-ionization effects  \citep[e.g.,][]{Rathore2015A} for ions in the CHIANTI package are important in flares. All of these limitations are being actively investigated with the intent of improving upon them in future generations of stellar flare model grids.

\begin{deluxetable}{ll}
\tabletypesize{\scriptsize}
\tablewidth{0pt}
\tablecaption{Species excluded from the CHIANTI optically thin radiative loss function}
\tablehead{
\colhead{Type} &
\colhead{Species} } 
\startdata 
free-free (f-f) & H II, He II, He III, Ca II, Ca III, Mg II, Mg III, Fe II, Fe III, Si II, Si III, C II, Al II \\
free-bound (f-b) & H I, He I, He II,  Mg I, Mg II,  Si I, Si II, C I, C II, Al I \\
bound-bound (b-b) & H I, He I, He II, Ca II,Mg II,  Fe II, Si II, C I, C II, Al II, O I, S I \\
\enddata
\tablecomments{ The f-b species listed refer to the respective bound-states.  The f-b transitions in Ca I, Ca II, Fe I, Fe II, and Al II are not in the CHIANTI v8.0.1 database and are thus also excluded.  The b-b transitions in Ca I, Mg I, Fe I, Si I, and Cr II are not in the CHIANTI v8.0.1 database and are thus also excluded.  We note that in CHIANTI v10.0.1 (the most recent version at the time of writing), b-b transitions in Cr II contribute a significant amount to the radiative losses at $T=10^4$ K.  }
\end{deluxetable}\label{table:thinloss}

\subsection{Summary of the Grid} 

The model grid consists of 80 simulations that span a large range of heating rates and dynamical evolution.  We organize the models into several groups.  The \texttt{main} grouping consists of large low-energy cutoffs with a ramp up/down flux injection;  the \texttt{const} group consists of large low-energy cutoff models with a constant flux injection; the \texttt{Ec37} group consists of models with $E_c = 37$ keV and hard power-law indices \citep{Allred2005, Allred2006, Kowalski2015}, as inferred through CTTM modeling of hard X-rays of the 2002 July 23 X-class flare \citep{Holman2003, Ireland2013}; the \texttt{sol} group consists of models with smaller low-energy cutoffs and softer (larger) power-law indices.  The \texttt{sol} group also includes a model with a longer, $\Delta t = 15$~s, injection of constant electron beam flux; this model is similar to the high-flux \texttt{c15s-5F11-25-4} flare model in solar gravity that has been analyzed in \citet{Kowalski2017Mar29, Kowalski2022}.  The \texttt{auxiliary} models include the following: a recalculation of a constant injection of F13 flux with $E_c=37$ keV and $\delta =3$ for 1.6~s \citep[following][]{Kowalski2016}, a high-flux model with a very hard power-law index, $\delta =2.5$ (\texttt{m2F12-37-2.5}), and a recalculation of the soft ($\delta = 7$), fully relativistic ($E_c = 500$ keV) electron beam model (\texttt{m2F12-500-7}) that was used in \cite{Kowalski2017Broadening} as a phenomenological model for a ``Vega-like" photospheric spectrum that was reported in the decay phase of a white-light megaflare \citep{Kowalski2013}.  Most models in the grid are calculated until $t = 10$~s, which allows the relaxation of each impulsively energized chromospheric region to be followed.

It is also customary to test electron beam models against flare models that include energy transport through thermal conduction, given an ad hoc energy deposition in the corona \citep{Fisher1989, Reep2016, Kowalski2017Mar29,Namekata2020}.  We do not specify the energy source, but it could plausibly represent heating across Petschek shocks in the reconnection process \citep{Longcope2016}.  Note, however, that gas compression and other physics of magnetic field retraction are obviously not included in the \texttt{RADYN} simulations, which begin with a static, semi-circular loop.  For the \texttt{thermal} models, flare heating is simulated as an ad hoc energy deposition  of 125 erg cm$^{-3}$ s$^{-1}$ at $z > 7.5$ Mm.  One model applies this coronal heating for $\Delta t = 2.3$~s, and another for $\Delta t = 10$~s.  

We include five models from the \citet{Namekata2020} grid of models in a separate \texttt{N+20} grouping; these models were calculated with the same version of the \texttt{RADYN} code.  Three triangular (\texttt{t}) pulses have a  maximum beam flux injection at $t=2$~s.  These are the \texttt{tF12-37-3}, \texttt{tF12-37-5}, and \texttt{tF10-37-5} models.  \citet{Namekata2020} also studied the response to triangular, thermal pulses with peak heating rates at $t=8$~s at the coronal apex; the model identifications are \texttt{thermal-tF12} and \texttt{thermal-t5F10} with spatial integrations of the coronal heating rates that correspond to $\approx 10^{12}$ and $5\times 10^{10}$ \fnum, respectively. We refer the reader to \citet{Namekata2020} for further details. Several \texttt{N+20} models are somewhat redundant with models in other groupings.  For example, the \texttt{tF12-37-3} and the \texttt{mF12-37-3}  models are nearly the same except for a  $\Delta t = 1$~s  between the times of the peak injected fluxes. These two models highlight uncertainties that are linked to small variations in the assumptions about the injected beam flux time profiles.  The \texttt{thermal-tF12} has a maximum heating rate of $\approx 1200$ erg cm$^{-3}$ s$^{-1}$, which is one order of magnitude larger than the three other thermal heating models.  This model produces the hottest corona achieved among all models (Section \ref{sec:quantities}). Thus, there is value in including these five representative models from the \citet{Namekata2020} grid.

\section{Model Grid Analysis} \label{sec:analysis}

We divide the model analysis into three parts.  First, we summarize the calculated quantities from the optical and near-ultraviolet grid spectra in Section \ref{sec:quantities}.  From these calculated quantities and the detailed emission line spectra, we demonstrate that the deep heating predictions of the model grid explain many of the observational constraints in the literature (Sections \ref{sec:global} - \ref{sec:detailed}).

\subsection{Calculated Spectral Quantities} \label{sec:quantities}

We follow \citet{Osten2016}, \citet{Kowalski2017Broadening}, and \citet{Kowalski2022Frontiers} and calculate the temporally averaged radiative surface flux spectrum, $F_{\lambda}$, from each model\footnote{In \cite{Kowalski2022Frontiers}, the model surface flux spectra were abbreviated as $S_{\lambda}$; here, we do not want to confuse this with the source function, $S_{\nu}$.}.   Temporal averaging is a simple method to emulate many heating and cooling loops at different stages in their impulsive evolution on the star in a manner analogous to multi-thread modeling of solar flares \citep{Hori1997, Reeves2002, Warren2006}.  The interpretation of $F_{\lambda}$ is that every model snapshot has an equal filling factor (area) in the flaring source over an exposure time, and all loops in the source are subject to the same heating pulse with uniformly distributed lags \citep[\emph{cf.} the superposition principle;][]{AschwandenAlexander2001}.  These are sufficient assumptions to first order, and we expand upon this interpretation in Section \ref{sec:detailed}.

The \texttt{main} group of models is averaged over $t = 0 - 10$~s and the \texttt{const} group is averaged over $t = 0 -5$~s.  The \texttt{thermal} models are averaged over $\Delta t = 10$~s, beginning when the conductive pulses reach the chromosphere. The flare-only, time-averaged radiative surface flux spectra are denoted as $F_{\lambda}^{^\prime}$, and the flare-only radiative surface flux spectra at time $t$ are indicated as $F_{\lambda}^{\prime}(t)$.  Note that Figure \ref{fig:ContSpec} displays $F_{\lambda}(t=1\rm{s})$ for the \texttt{mF13-85-3} and \texttt{m5F11-25-4} beam models.  

From the model spectra, we calculate several quantities to directly compare to stellar flare observations (Section \ref{sec:global}).  Table \ref{table:calculations} lists several of the  calculations, and others are described in the supplemental document that is referenced in the Appendix.  The columns of Table \ref{table:calculations} are the following:

\begin{enumerate}
\item Unique model identifier.

\item C4170$^{\prime}$:  The value of $F_{\lambda}^{^\prime}$ at the continuum wavelength of $\lambda=4170$ \AA. Note that some small values are negative.  The negative continuum fluxes can result from weak heating rates, which elevate the collisional rates in the chromosphere  enough to diminish some of the background photospheric radiation \citep[e.g.,][]{Allred2005, Kowalski2017Mar29}.

\item $F_{\rm{H}\gamma}^{\prime}$:  continuum-subtracted $F_{\lambda}^{^\prime}$ integrated over the wavelengths, $\lambda_1$ to $\lambda_2$, of the Balmer H$\gamma$ line; this is calculated from the temporally averaged surface flux spectrum (e.g., from $t=0-10$~s), and the pre-flare value (in the first row)  calculated from $F_{\lambda}(t=0)$  has been subtracted.

\item $F_{\rm{H}\gamma}^{\prime}$ /C4170$^{\prime}$:  the ratio of column (3) to (2)

\item $T_{\rm{BB}}(1\rm{s})$:  blackbody temperature fit at the model continuum wavelengths over $\lambda = 4000-4800$ \AA, following \citet{Kowalski2013};  this is calculated from $F_{\lambda}^{\prime}(t=1\rm{s})$.

\item $T_{\rm{BB}}(2.2\rm{s})$:  blackbody temperature fit at the model continuum wavelengths over $\lambda = 4000-4800$ \AA, following \citet{Kowalski2013};  this is calculated from $F_{\lambda}^{\prime}(t=2.2\rm{s})$.

\item $T_{\rm{BB}}$:  blackbody temperature fit at the model continuum wavelengths over $\lambda = 4000-4800$ \AA, following \citet{Kowalski2013};  this is calculated as $F_{\lambda}^{\prime}$ from the time-averaged surface flux spectrum.  The temporal calculations of $T_{\rm{BB}}$ (cols 5 and 6) give a sense by which the variation of the optical color temperatures differ from the color temperature of the average spectrum (col 7).
 A similar blackbody color temperature, $T_{\rm{FcolorR}}$ (not listed in this table), is calculated from the ratio of $F_{\lambda}^{\prime}$ at $\lambda = 4170$ \AA\ and $\lambda = 6010$ \AA. Spectra indicate that $T_{\rm{BB}} \ne T_{\rm{FcolorR}}$ \citep{Kowalski2016}.  In Section \ref{sec:global}, the model values of $T_{\rm{FcolorR}}$ are compared to comprehensive observational constraints over a broader wavelength range than fit with $T_{\rm{BB}}$.

\item C3615$^{\prime}$/C4170$^{\prime}$:  the Balmer jump ratio, which is the flare-only continuum flux at $\lambda=3615$ \AA\ divided by column 2;  this is calculated from $F_{\lambda}^{\prime}$.

\item H$\gamma$ Eff. Width:  the effective width \citep{Kowalski2022} of the H$\gamma$ line; this is calculated from $F_{\lambda}$ after a linear fit to the nearby continuum flux on both sides of the emission line has been interpolated and subtracted:

\begin{equation} \label{eq:effective_width}
\Delta \lambda_{\rm{eff}} = \int_{\lambda_1}^{\lambda_2} \frac{F_{\lambda}(\lambda) - F_{\lambda,\rm{cont}}(\lambda)}{F_{\lambda}(\lambda_{\rm{max}})- F_{\lambda,\rm{cont}}(\lambda_{\rm{max}})  } d\lambda 
\end{equation}

\noindent where $\lambda_1 = 4312$ \AA\ and $\lambda_2 = 4368$ \AA.

\item $T_{\rm{plasma,max}}$:  the maximum value of the plasma temperature, considering all times in the model.  The time at which the maximum plasma temperature occurs is $t_{\rm{max}}$.

\item $T_{\rm{XEUV,max}}$:  the XEUV continuum thermal bremsstrahlung color temperature calculated from the atmospheric snapshot at the same time as in column 10. This is calculated from $F_{\lambda}(t=t_{\rm{max}})$. We integrate the emissivity from hydrogen and helium at plasma temperatures $T > 1$ MK.  The  color temperature is calculated from the ratio of the emergent optically thin flux at $\lambda = 8$ \AA\ to that at $\lambda = 35$ \AA.  An X-ray color temperature ($T_{\rm{X,max}}$; not listed) is calculated from the ratio of the emergent flux at $\lambda = 0.5$ \AA\ to that at $\lambda = 4$ \AA.
\end{enumerate}

The values in Table \ref{table:calculations} serve as a comprehensive overview of the observables from the models.  Before turning to comparisons to observations of optical flare properties, a few comments about the coronal predictions (last two columns) are warranted.  First, many of the models with large low-energy cutoffs do not generate enough Coulomb heating in the corona to produce a response at plasma temperatures $T > 1$ MK.  This is certainly inconsistent with many stellar flare observations, which suggest temperatures in the range around $T \approx 15$ MK to $T \gg 50$ MK \citep[e.g.,][]{Gudel2004Rev, Mullan2006, Liefke2010, Osten2005, Osten2010, Osten2016}.  Second, a few of the models produce maximum plasma temperatures that are extremely high, $T_{\rm{plasma,max}} \approx 40-90$ MK.  However, the thermal bremsstrahlung temperatures that are calculated from XEUV continuum spectra over  $\lambda = 8-35$ \AA\ (last column) are much cooler.  This systematic discrepancy occurs  because the emergent radiation at these wavelengths originates mostly from the cooler, denser material at the base of the evaporation flows.  The X-ray color temperatures ($T_{\rm{X,max}}$; not listed) from $0.5-4$ \AA\ are much closer to the values of the maximum plasma temperatures, though in the hottest model flare coronae, there are still differences of $T_{\rm{X,max}} - T_{\rm{plasma,max}} \approx 5 - 20$ MK.

The largest coronal plasma temperatures in our models are attained when the evaporation shock fronts reach the looptop.  For example, in the \texttt{mF13-37-3} model, a $v_z \approx 1750 $ km s$^{-1}$ evaporation reaches the looptop at $t = 6$~s, and then there is a jump in the apex temperature from $T \approx 50$ MK to $\approx 90$ MK by $t=7$~s. The temperature explosions in the high corona  in our models are facilitated by a reflecting upper boundary condition (which we chose to emulate a symmetrical flare loop).  We analyze the terms that contribute to changes of the internal energy per gram, and we find that the rapid temperature increase is due to viscosity and compression in the large velocity gradient that occurs as flows from both sides of the loop collide.  Notably, these effects have been previously discussed in detail in the high-flux hydrodynamic model of the coronal explosion of CN Leo \citep{Schmitt2008}. We also are able to follow the \texttt{mF13-37-3} model to $t=20$~s.  We find that the evolution of the coronal densities and velocities is qualitatively very similar to Fig.\ 3 of \citet{Reeves2007} on much longer timescales over much larger loop lengths.  The main difference is that a secondary chromospheric evaporation wave in our model begins at $t \approx 10$~s due to the  looptop heat conduction pulse traveling down the loop ahead of the flows from the other side.  The secondary evaporation and the flow from the other side collide in the middle corona by about $t=14$~s.   By $t=20$~s, most of the corona is still very dense, $n_e = 2.5 - 5\times10^{11}$ cm$^{-3}$ and very hot, $T \approx 40$ MK.  It is reassuring that three entirely different numerical codes predict gross commonalities in the coronal evolution driven by large heating rates.   

The impulsively-heated \texttt{RADYN} models in our grid are, however, not intended to include the late evolution of the corona long after the end of electron-beam energy injection, which we assume is very short within each loop in stellar flares (see Section \ref{sec:discussion} for further discussion).  The optical and NUV radiation decrease to very faint levels by $t \approx 10$~s (Figure \ref{fig:Figure_BeamEvol}), at which point the integrated coronal emission is still increasing in some models. Only a few of the models in our grid are extended to $t > 10$~s, but all are averaged over a short duration ($\Delta t = 5$ or $10$~s) to provide the average quantities in Table \ref{table:calculations}. The cooling over the late evolution of the coronae may be further affected by transport physics \citep{Bian2018, Zhu2018, Allred2022, Ashfield2023} that are not considered in this generation of models.

\startlongtable
\begin{deluxetable}{ccccccccccc}
\centerwidetable
\tablehead{ \colhead{Model ID} & \colhead{C4170$^{\prime}$} & \colhead{$F_{\rm{H}\gamma}^{\prime}$ } &\colhead{$F_{\rm{H}\gamma}^{\prime}$ /C4170$^{\prime}$} & \colhead{$T_{\rm{BB}}(1\rm{s})$} & \colhead{$T_{\rm{BB}}(2.2\rm{s})$} & \colhead{$T_{\rm{BB}}$} & \colhead{C3615$^{\prime}$/C4170$^{\prime}$} & \colhead{H$\gamma$ Eff. Width} & \colhead{$T_{\rm{plasma,max}}$}& \colhead{$T_{\rm{XEUV,max}}$} \\ 
(1) & (2) & (3) & (4) & (5) & (6) & (7) & (8) & (9) & (10) & (11) } 
\startdata 
         & erg cm$^{-2}$ s$^{-1}$ \AA$^{-1}$  & erg cm$^{-2}$ s$^{-1}$ & \AA & K & K & K & \nodata & \AA & $10^6$ K & $10^6$ K  \\
\hline
\hline
$t = 0$~s   & 2.5e+05 &  6.24e+06 &  25 & \nodata & \nodata & 4490 & 0.7 & 0.49 & 5.0 & 4.2 \\
\hline
main group &           &           &      &        &         &         &      &       &     &     \\
\hline
mF10-85-3 & 1.54e+04 & 1.07e+07 & 691.2 & 6200 & 5400 & 5600 & 40.62 & 1.06 & 5.0 & 4.2\\
mF10-150-3 & 8.46e+03 & 6.85e+06 & 810.4 & 4500 & 6000 & 6600 & 56.57 & 0.89 & 5.0 & 4.2 \\
mF10-200-3 & 1.28e+03 & 4.98e+06 & 3885.9 & $^{\dagger}$\nodata & \nodata & \nodata & 309.23 & 0.78 & 5.0 & 4.2 \\
mF10-350-3 & -1.86e+04 & 2.31e+06 & -124.6 & $^{\dagger}$\nodata & \nodata & \nodata & -15.25 & 0.62 & 5.0 & 4.2 \\
mF10-500-3 & -3.03e+04 & 1.48e+06 & -48.7 & $^{\dagger}$\nodata & \nodata & \nodata & -8.17 & 0.57 & 5.0 & 4.2 \\
mF11-85-3 & 4.78e+05 & 8.86e+07 & 185.2 & 5200 & 5200 & 5100 & 12.02 & 3.84 & 5.0 & 4.3 \\
mF11-150-3 & 5.39e+05 & 6.93e+07 & 128.7 & 5300 & 5300 & 5200 & 9.32 & 4.27 & 5.0 & 4.2 \\
mF11-200-3 & 5.61e+05 & 6.00e+07 & 107.0 & 5400 & 5400 & 5200 & 7.99 & 4.11 & 5.0 & 4.2 \\
mF11-350-3 & 1.20e+06 & 9.09e+07 & 75.5 & 5300 & 5400 & 5400 & 6.07 & 5.66 & 5.0 & 4.2 \\
mF11-500-3 & 5.25e+05 & 2.98e+07 & 56.8 & 5000 & 5400 & 5200 & 4.84 & 2.53 & 5.0  & 4.2 \\
mF12-85-3 & 8.33e+06 & 7.53e+08 & 90.4 & 6100 & 6200 & 5900 & 4.97 & 6.55 & 5.8 & 5.1 \\
mF12-150-3 & 1.37e+07 & 6.58e+08 & 48.1 & 6400 & 6600 & 6300 & 3.45 & 11.08 & 5.0 & 4.3 \\
mF12-200-3 & 1.61e+07 & 4.85e+08 & 30.1 & 6600 & 7000 & 6500 & 2.63 & 14.41 & 5.0 & 4.3 \\
mF12-350-3 & 1.89e+07 & 2.35e+08 & 12.4 & 7000 & 7900 & 7100 & 1.71 & 23.97 & 5.0 & 4.2\\
mF12-500-3 & 2.00e+07 & 1.31e+08 & 6.5 & 7200 & 8500 & 7400 & 1.41 & 26.05 & 5.0 & 4.2 \\
m2F12-85-3 & 2.00e+07 & 1.15e+09 & 57.3 & 6700 & 7100 & 6600 & 3.38 & 8.05 & 7.2 & 5.9 \\
m2F12-150-3 & 3.72e+07 & 9.46e+08 & 25.5 & 7600 & 8800 & 7700 & 2.28 & 11.76 & 5.1 & 4.4 \\
m2F12-200-3 & 4.55e+07 & 5.87e+08 & 12.9 & 8300 & 10000 & 8500 & 1.79 & 12.16 & 5.0 & 4.2  \\
m2F12-350-3 & 7.71e+07 & -2.60e+08 & -3.4 & 10000 & 12500 & 10400 & 1.06 & 6.62 & 5.0 & 4.2 \\
m2F12-500-3 & 5.80e+07 & -4.10e+08 & -7.1 & 10600 & 13600 & 11100 & 0.86 & 14.95 & 5.0 & 4.2 \\
mF13-85-3 & 1.14e+08 & 1.62e+09 & 14.2 & 11100 & 12300 & 10800 & 1.73 & 9.13 & 16.1 & 10.1 \\
mF13-150-3 & 1.54e+08 & 1.16e+09 & 7.5 & 12900 & 13400 & 12100 & 1.54 & 10.33 & 7.0 & 5.7 \\
mF13-200-3 & 1.73e+08 & 6.21e+08 & 3.6 & 13800 & 14200 & 12800 & 1.42 & 10.38 & 5.3 & 4.7 \\
mF13-350-3 & 1.93e+08 & -7.08e+08 & -3.7 & 16200 & 16500 & 14900 & 1.12 & 13.78 & 5.0 & 4.2 \\
mF13-500-3 & 1.87e+08 & -1.37e+09 & -7.3 & 18200 & 18900 & 16500 & 0.95 & 14.96 & 5.0 & 4.2 \\
\hline
const group &           &           &      &        &         &         &      &       &     &     \\
\hline
cF11-85-3 & 8.40e+05 & 1.48e+08 & 176.4 & 5200 & 5300 & 5200 & 11.56 & 4.92 & 5.0 & 4.3 \\
cF11-150-3 & 9.35e+05 & 1.15e+08 & 123.4 & 5300 & 5400 & 5300 & 9.09 & 5.97 & 5.0 & 4.2 \\
cF11-200-3 & 9.68e+05 & 1.00e+08 & 103.6 & 5400 & 5400 & 5300 & 7.84 & 6.08 & 5.0 & 4.2 \\
cF11-350-3 & 9.69e+05 & 6.94e+07 & 71.5 & 5400 & 5500 & 5400 & 5.75 & 5.23 & 5.0 & 4.2 \\
cF11-500-3 & 8.91e+05 & 4.95e+07 & 55.6 & 5400 & 5500 & 5400 & 4.73 & 4.00 & 5.0 & 4.2 \\
c5F11-85-3 & 6.32e+06 & 7.15e+08 & 113.2 & 5900 & 5900 & 5800 & 6.11 & 6.14 & 5.3 & 4.6 \\
c5F11-150-3 & 9.16e+06 & 5.76e+08 & 62.8 & 5900 & 6100 & 5800 & 4.55 & 10.80 & 5.0 & 4.2 \\
c5F11-200-3 & 9.98e+06 & 4.64e+08 & 46.5 & 6000 & 6200 & 5900 & 3.72 & 17.07 & 5.0 & 4.2 \\
c5F11-350-3 & 1.10e+07 & 3.35e+08 & 30.6 & 6200 & 6500 & 6100 & 2.69 & 20.95 & 5.0 & 4.2 \\
c5F11-500-3 & 1.13e+07 & 2.63e+08 & 23.4 & 6300 & 6700 & 6300 & 2.27 & 21.34 & 5.0 & 4.2 \\
cF12-85-3 & 1.45e+07 & 1.14e+09 & 79.1 & 6200 & 6400 & 6200 & 4.30 & 7.86 & 5.8 & 4.9 \\
cF12-150-3 & 2.61e+07 & 1.12e+09 & 42.8 & 6600 & 7200 & 6600 & 2.99 & 11.80 & 5.0 & 4.3 \\
cF12-200-3 & 3.13e+07 & 7.64e+08 & 24.4 & 6900 & 7800 & 7000 & 2.26 & 14.74 & 5.0 & 4.3 \\
cF12-350-3 & 3.75e+07 & 2.22e+08 & 5.9 & 7600 & 9200 & 7800 & 1.39 & 37.62 & 5.0 & 4.2\\
cF12-500-3 & 3.90e+07 & 4.21e+07 & 1.1 & 8000 & 10000 & 8200 & 1.16 & 23.82 & 5.0 & 4.2 \\
c2F12-85-3 & 3.50e+07 & 1.58e+09 & 45.1 & 7000 & 7700 & 7000 & 2.81 & 10.13 & 7.2 & 5.8 \\
c2F12-150-3 & 6.54e+07 & 1.39e+09 & 21.3 & 8300 & 9700 & 8400 & 2.03 & 12.41 & 5.1 & 4.4 \\
c2F12-200-3 & 8.05e+07 & 8.87e+08 & 11.0 & 9200 & 10800 & 9300 & 1.68 & 13.02 & 5.0 & 4.2 \\
c2F12-500-3 & 1.01e+08 & -8.44e+08 & -8.3 & 12300 & 15100 & 12000 & 0.81 & 14.14 & 5.0 & 4.2 \\
c5F12-85-3 & 9.78e+07 & 1.65e+09 & 16.9 & 9600 & 11000 & 9600 & 1.81 & 11.42 & 11.5 & 7.6 \\
c5F12-150-3 & 1.46e+08 & 1.20e+09 & 8.2 & 11500 & 12600 & 11300 & 1.55 & 12.17 & 5.6 & 4.8\\
c5F12-200-3 & 1.69e+08 & 7.65e+08 & 4.5 & 12500 & 13300 & 12100 & 1.44 & 12.18 & 5.1 & 4.4 \\
cF13-85-3 & 1.65e+08 & 1.44e+09 & 8.7 & 11900 & 12800 & 11600 & 1.54 & 11.48 & 17.3 & 8.9 \\
cF13-150-3 & 2.10e+08 & 8.73e+08 & 4.2 & 13500 & 14100 & 13200 & 1.42 & 10.38 & 6.9 & 5.6 \\
cF13-200-3 & 2.27e+08 & 5.07e+08 & 2.2 & 14300 & 14900 & 14000 & 1.37 & 9.38 & 5.3 & 4.6 \\
cF13-350-3 & 2.43e+08 & -5.27e+08 & -2.2 & 16500 & 17000 & 15800 & 1.23 & 9.02 & 5.0 & 4.2 \\
cF13-500-3 & 2.64e+08 & -1.19e+09 & -4.5 & 18600 & 19100 & 17500 & 1.14 & 10.12 & 5.0 & 4.2 \\
\hline
Ec37 group &           &           &      &        &         &         &      &       &     &     \\
\hline
mF10-37-3 & 1.69e+04 & 1.64e+07 & 967.6 & 5100 & 5100 & 5200 & 41.57 & 1.15 & 5.1 & 4.2 \\
m5F10-37-3 & 1.49e+05 & 7.50e+07 & 502.3 & 5600 & 5400 & 5300 & 17.30 & 1.71 & 5.4 & 4.5 \\
mF11-37-3 & 3.21e+05 & 1.24e+08 & 385.6 & 5700 & 5500 & 5400 & 13.33 & 1.95 & 5.8 & 4.9\\
m2F11-37-3 & 6.55e+05 & 1.93e+08 & 294.2 & 5700 & 5500 & 5500 & 10.88 & 2.31 & 6.7 & 5.6 \\
m5F11-37-3 & 1.67e+06 & 3.41e+08 & 203.8 & 5800 & 5700 & 5600 & 8.47 & 3.05 & 9.4 & 7.2\\
mF12-37-3 & 3.51e+06 & 5.45e+08 & 155.2 & 5900 & 5800 & 5700 & 6.87 & 3.85 & 13.6 & 8.9  \\
m2F12-37-3 & 7.93e+06 & 9.33e+08 & 117.8 & 6100 & 6100 & 5900 & 5.33 & 4.99 & 21.1 & 10.3 \\
m5F12-37-3 & 2.92e+07 & 1.99e+09 & 68.1 & 6700 & 7400 & 6800 & 3.22 & 8.41 & 39.3 & 15.9 \\
m7.5F12-37-3 & 5.34e+07 & 2.37e+09 & 44.4 & 7400 & 8700 & 7900 & 2.44 & 10.59 & 56.1 & 18.1 \\
mF13-37-3 & 7.94e+07 & 2.22e+09 & 27.9 & 8200 & 10500 & 9500 & 1.96 & 12.11 & 88.2 & 17.8 \\
m5F10-37-5 & 1.37e+05 & 8.54e+07 & 622.9 & 6000 & 5700 & 5600 & 14.03 & 1.47 & 5.5 & 4.7  \\
mF12-37-5 & 1.10e+06 & 4.14e+08 & 375.9 & 6200 & 6100 & 6000 & 9.31 & 2.50 & 17.0 & 9.3 \\
\hline
sol group &           &           &      &        &         &         &      &       &     &     \\
\hline
mF10-17-3 & 1.60e+04 & 2.47e+07 & 1542.0 & 5400 & 5300 & 5300 & 39.73 & 0.97 & 5.4 & 4.5 \\
mF11-17-3 & 1.58e+05 & 9.39e+07 & 595.7 & 5800 & 5500 & 5500 & 16.27 & 1.51 & 9.7 & 7.3 \\
m2F11-17-3 & 3.37e+05 & 1.56e+08 & 461.9 & 5800 & 5500 & 5500 & 13.15 & 1.84 & 14.1 & 9.1 \\
m5F10-17-5 & 3.95e+04 & 4.67e+07 & 1182.4 & 5900 & 5600 & 5400 & 19.00 & 1.04 & 9.1 & 6.9 \\
m5F11-25-4 & 6.35e+05 & 2.95e+08 & 464.6 & 6000 & 5900 & 5800 & 10.80 & 2.20 & 18.6 & 10.0 \\
c15s-5F11-25-4 & 3.30e+06 & 1.33e+09 & 402.4 & 6000 & 6000 & 5700 & 9.12 & 5.33 & 60.3 & 17.9 \\
\hline
thermal group &           &           &      &        &         &         &      &       &     &     \\
\hline
thermal-long & 3.17e+04 & 6.62e+07 & 2086.2 & 4400 & 4400 & 5300 & 18.94 & 1.28 & 31.9 & 10.4 \\
thermal-short & 1.36e+03 & 1.31e+07 & 9652.2 & 5700 & 5700 & 4600 & 175.70 & 0.70 & 21.4 & 9.0 \\
\hline
auxiliary group &           &           &      &        &         &         &      &       &     &     \\
\hline
m2F12-37-2.5 & 1.72e+07 & 8.65e+08 & 50.2 & 6500 & 7200 & 6500 & 2.99 & 5.80 & 15.8 & 9.5 \\
c1.6s-F13-37-3 & 7.40e+07 & 2.31e+09 & 31.3 & 9300 & 8300 & 9000 & 2.13 & 12.02 & 71.4 & 17.9 \\
c2F12-500-7 & 9.76e+07 & -8.25e+08 & -8.5 & 13600 & 14800 & 12200 & 0.91 & 13.23 & 5.0 & 4.2\\
\hline
N+20 group &           &           &      &        &         &         &      &       &     &     \\
\hline
tF12-37-3 & 2.41e+06 & 3.64e+08 & 151.0 & 5800 & 5900 & 5700 & 6.85 & 4.02 & 12.3 & 8.0 \\
tF12-37-5 & 6.50e+05 & 2.32e+08 & 357.3 & 6400 & 6100 & 6000 & 9.55 & 2.46 & 16.5 & 8.8\\
tF10-37-5 & 3.07e+04 & 2.19e+07 & 713.6 & 4900 & 5000 & 4900 & 24.22 & 1.25 & 5.0 & 4.2 \\
thermal-t5F10 & -5.60e+03 & 6.76e+06 & -1209.0 & 5200 & 5200 & 5400 & -35.76 & 0.64 & 16.0 & 7.9 \\
thermal-tF12 & 1.48e+05 & 1.52e+08 & 1028.5 & 4400 & 5100 & 5500 & 14.53 & 2.01 & 89.9 & 9.7 \\
\enddata \label{table:calculations}
\tablenotetext{\dagger}{The mF10-200-3 model has very faint optical continuum flux (column 2), and thus reliable blackbody fits (columns 5, 6, and 7) can not be determined. The heating in the mF10-350-3 and mF10-500-3 models result in an attenuation of the pre-flare photospheric flux, thus causing negative continuum fluxes and unreliable blackbody fits as well.} 
\end{deluxetable}

\subsection{Comparison to Global Trends}  \label{sec:global}

\begin{figure*}
\gridline{\fig{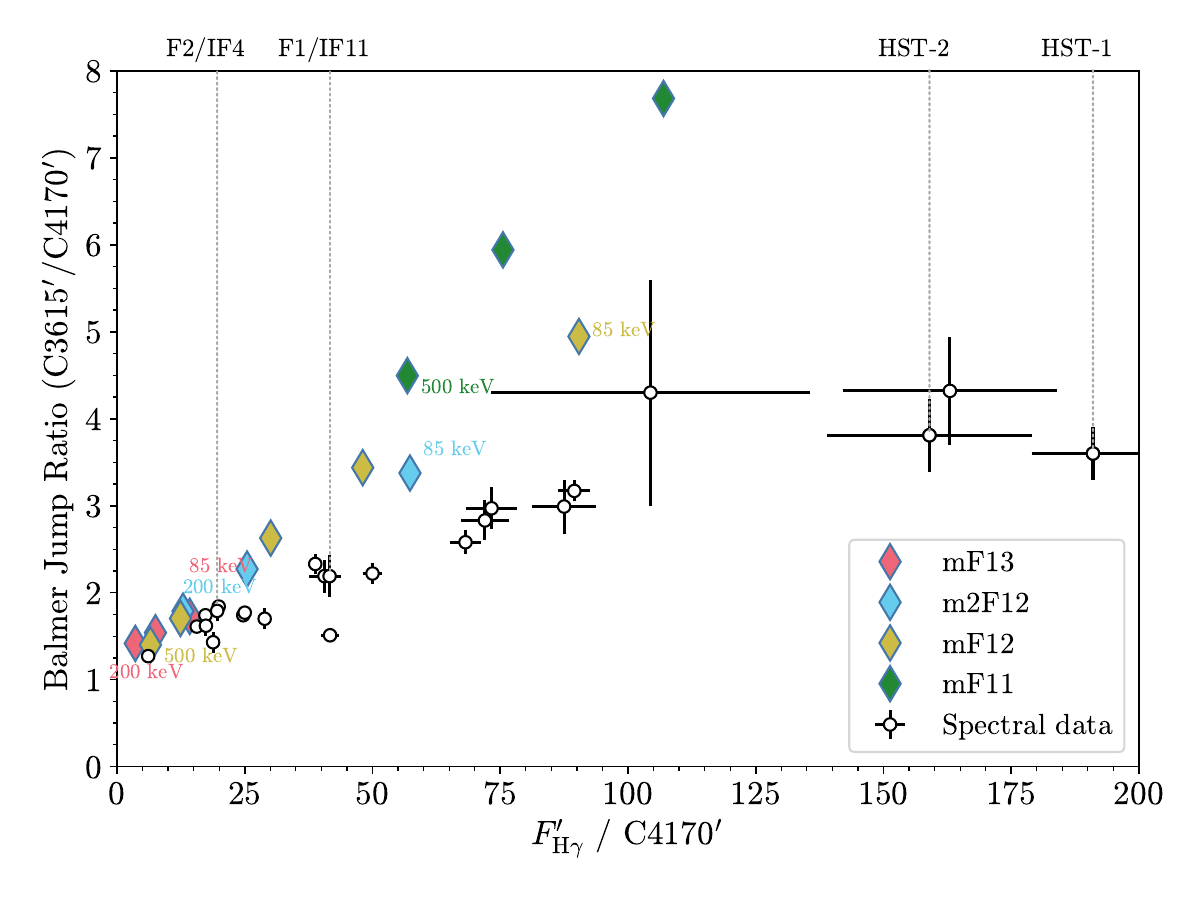}{0.5\textwidth}{(a)}
          \fig{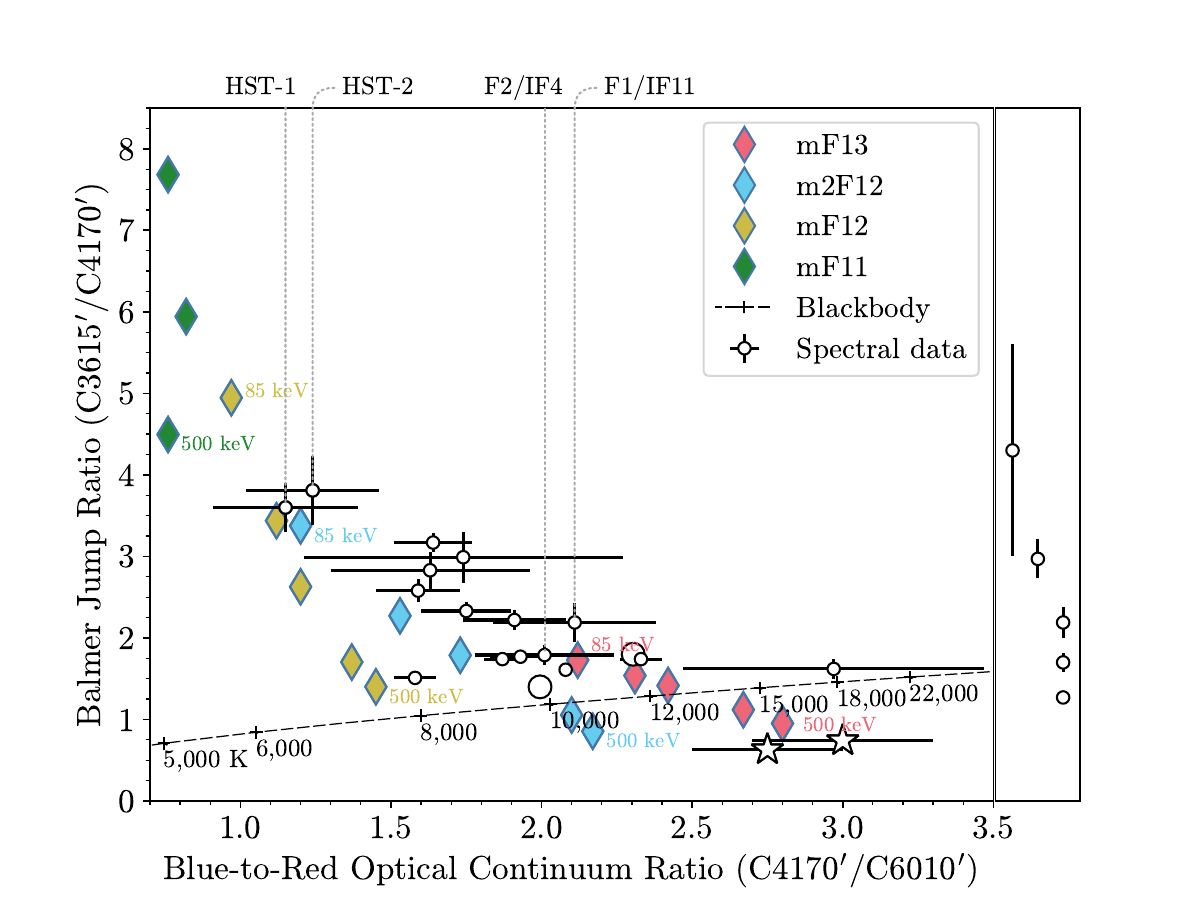}{0.5\textwidth}{(b)}
          }
\gridline{\fig{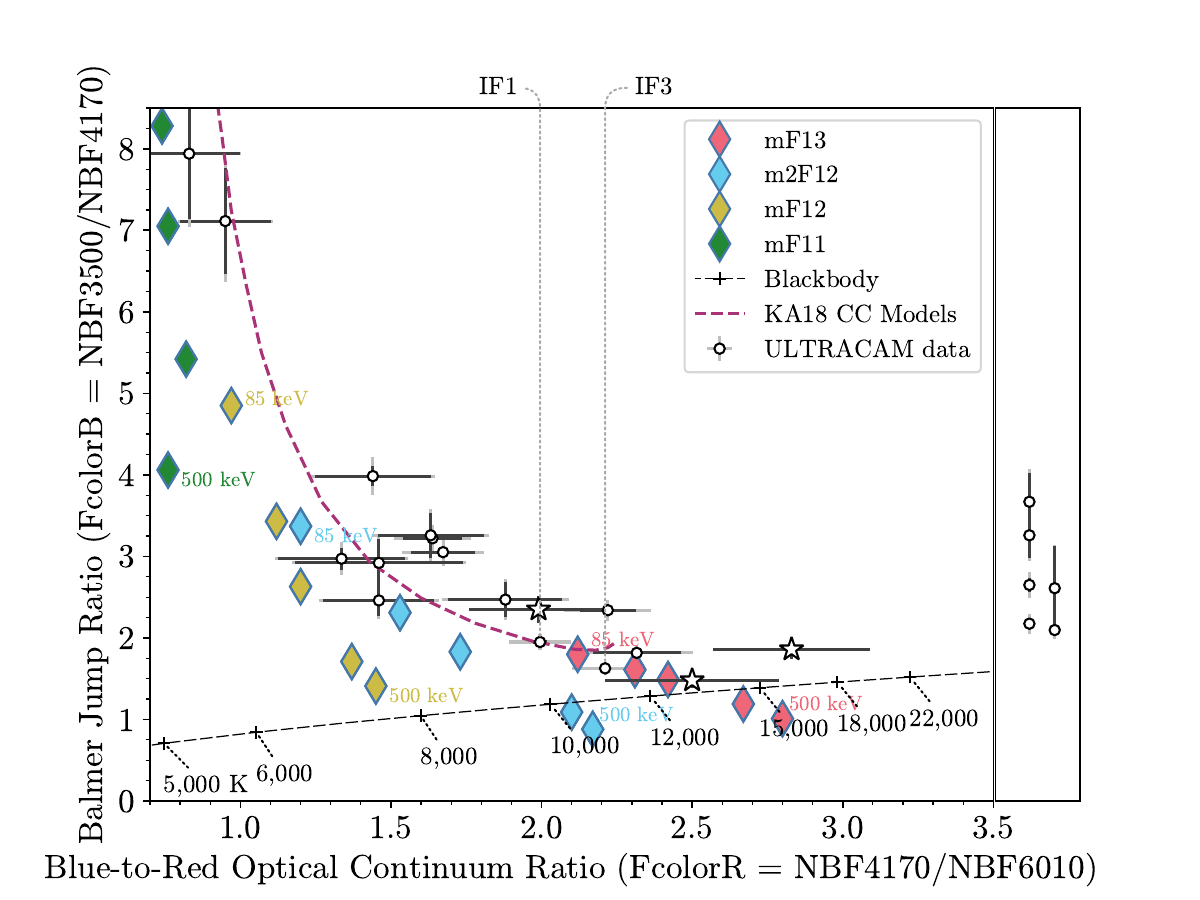}{0.75\textwidth}{(c)} }
\caption{  The predictions from the \texttt{main} group of models with large values of $E_c$ and a range of nonthermal energy fluxes are compared to observational constraints.  (a) Balmer H$\gamma$ line-to-continuum ratios over the time corresponding to the optical continuum peak for a sample of flares with spectra.  (b) Color-color diagram for these flares.  (c) Color-color diagram for the peak phases of the flares with ULTRACAM data.  The ULTRACAM peak-phase colors are annotated for the large events that are referred to as ``impulsive flare 1'' (IF1) and ``impulsive flare 3'' (IF3) in \citet{Kowalski2016}.  The HST-1 and HST-2 data are added from \citet{Kowalski2019HST}, the F1/IF11 and F2/IF4 flares are added from \citet{Kowalski2013}, and the rest of the spectral data in (a) and (b) are obtained from \citet{Kowalski2013} and \citet{HP91}.  The open star symbols are calculated from ``newly-formed'' flare flux spectra during secondary events, and the large open circles are obtained from spectra during the impulsive phase of the Great Flare of AD Leo \citep{HP91}.  Blackbody color temperatures are indicated in (b) and (c), and the parameterized chromospheric condensation (CC) models from \citet{KA18} are included in (c). Error bars on the data indicate marginal 68\% uncertainties.  In panel (c), the systematic uncertainties are included in the gray error bars, while the black error bars are the statistical uncertainties only (note that the statistical uncertainties are very small for the IF1 and IF3 events).  In (b) and (c), flares with only Balmer jump calculations are shown in the right panels in the margins.
\label{fig:ColorColor}}
\end{figure*}

\citet{Kowalski2013} presented global trends in optical spectroscopic observations of dMe flares.  The continuum constraints are generally interpreted as evidence for white-light radiating sources that cannot be  explained by optically thin thermal recombination radiation or by a superhot free-free source \citep[see also][]{HF92}.
The spectra include the Balmer jump and extend to $\lambda \approx 9000$ \AA\ in the largest events.  The trends have been supplemented with spectra of smaller dMe events in \citet{Kowalski2016} and \citet{Kowalski2019HST}, and with high-time resolution ULTRACAM \citep{Dhillon2007} photometry in \citet{Kowalski2016}.  Figures \ref{fig:ColorColor}(a-c) compare several calculated quantities (Table \ref{table:calculations}) from the \texttt{main} group of models to the impulsive phase observations from these studies.  

The Balmer jump ratio correlates with the H$\gamma$ line-to-continuum ratio in both the models and the observations (Figure \ref{fig:ColorColor}(a)), but the models predict steeper slopes.  The \texttt{mF12} and \texttt{m2F12} models satisfactorily reproduce some of the smallest Balmer jumps, which tend to occur in the observations at the peaks of the most impulsive flares. However, these beam fluxes do not account for the relationship between the small Balmer jumps and the optical color temperatures of $T \approx 10,000$ K.  In Figure \ref{fig:ColorColor}(b), the Balmer jump ratios are shown against the optical color temperatures, $T_{\rm{FcolorR}}$, that are calculated from the ratio of the flare-only continuum fluxes at $\lambda=4170$ \AA\ and $\lambda = 6010$ \AA\ (Section \ref{sec:quantities}).  The \texttt{main} \texttt{mF13} models satisfactorily reproduce the range of the largest optical continuum flux ratios that extend to the flare spectra with the Balmer continuum in absorption \citep[open star symbols;][]{Kowalski2013}.  The large open circle at $(x,y) =  (2.3, 1.8)$ is calculated from the peak-phase spectrum\footnote{The blackbody fit to the continuum regions at $\lambda \approx 3915-4420$ \AA\ \citep{Kowalski2013} is extrapolated to $\lambda = 6010$ \AA.} of the Great Flare of AD Leo \citep{HP91}, which falls close to the prediction of the \texttt{mF13-150-3} model \citep{Kowalski2022Frontiers}.  

Figure \ref{fig:ColorColor}(c) contains additional constraints on Balmer jump ratios and optical blackbody color temperatures at the peak phases of a sample of flares with ULTRACAM data \citep{Kowalski2016}.  
The two largest events in this sample correspond to the IF1 and IF3 events, whose peak colors straddle the prediction of the \texttt{mF13-85-3} model \citep{Kowalski2023}.  Parameterized models of dense chromospheric condensations \citep[purple dashed line;][]{KA18} reproduce the general trend in the peak-flare data. However, the models at the bottom right of the track generate extremely broad Balmer emission line profiles that are inconsistent with observations \citep{Kowalski2017Broadening, Kowalski2022Frontiers}.  The chromospheric condensation models also do not extend to the lower right end of the observed continuum colors, which are accounted for by the \texttt{main} F13 models in the \texttt{RADYN} grid.  

In summary, the high-flux ($10^{12} - 10^{13}$ \fnum), large $E_c$ \texttt{main} models generate heating rates that reproduce the most extreme dMe flare continuum spectral properties.   Models 
with fluxes between $10^{10} - 10^{12}$ \fnum\ and smaller low-energy cutoffs between $17-37$ keV (e.g., the \texttt{mF12-37-3} and \texttt{m5F11-25-4} models; Table \ref{table:calculations}) produce Balmer jumps, optical color temperatures, and Balmer line-to-continuum ratios that are inconsistent with the cluster of flux ratios from impulsive-phase spectral observations.

\subsection{Hydrogen Broadening from Flare Models with Deep Heating} \label{sec:detailed}

The superposition of a lower-flux beam model with one of the large-$E_c$, high-flux beam models brings  many of the discrepancies (e.g., line-to-continuum ratios) further in line with the observations in Figure \ref{fig:ColorColor}.  A linear, least-squares regression analysis with our grid of models was presented in \cite{Kowalski2022Frontiers}, who fit the broadband FUV-to-optical continuum shapes and the broad hydrogen emission lines in the Great Flare of AD Leo \citep{HP91}.  In this section, we model the optical spectra of a remarkable white-light stellar flare observed at much higher resolving power, which facilitates rigorous tests of the predictions of the large optical depths and electron densities produced in the large-$E_c$, high-flux models. 

An energetic flare from the dM4.5e star YZ CMi was observed on 2008--Jan--21 with the 2.03 m Telescope Bernard Lyot at the Pic du Midi Observatory.  The cross-dispersed NARVAL spectropolarimeter \citep{Silvester2012} was employed with a 2.\arcsec8 slit and 1200~s exposure times.  The linear dispersion at $\lambda \approx 4300$ \AA\ is $0.027$ \AA\ pix$^{-1}$.  A series of four consecutive Stokes I spectra covering $\lambda = 3700$ \AA\ to $10,000$ \AA\ were obtained from 22:28 to 23:32 UT at an airmass around $1.3$.  The spectra are not normalized by the continuum, but they are divided by the large-scale sensitivity of a flat field.  The data were first presented in the context of other stellar flares with blueshifts in the Balmer lines \citep{Vida2019}.  We retrieved these  spectra from the PolarBase archive \citep{PolarBase1, PolarBase2}, and we uniformly applied a wavelength shift of $\Delta \lambda = -0.35$ \AA, which was required to align the pre-flare Ca II K emission line centroids in the model and observations.

\begin{figure}
	\begin{center}
\includegraphics[scale=.75]{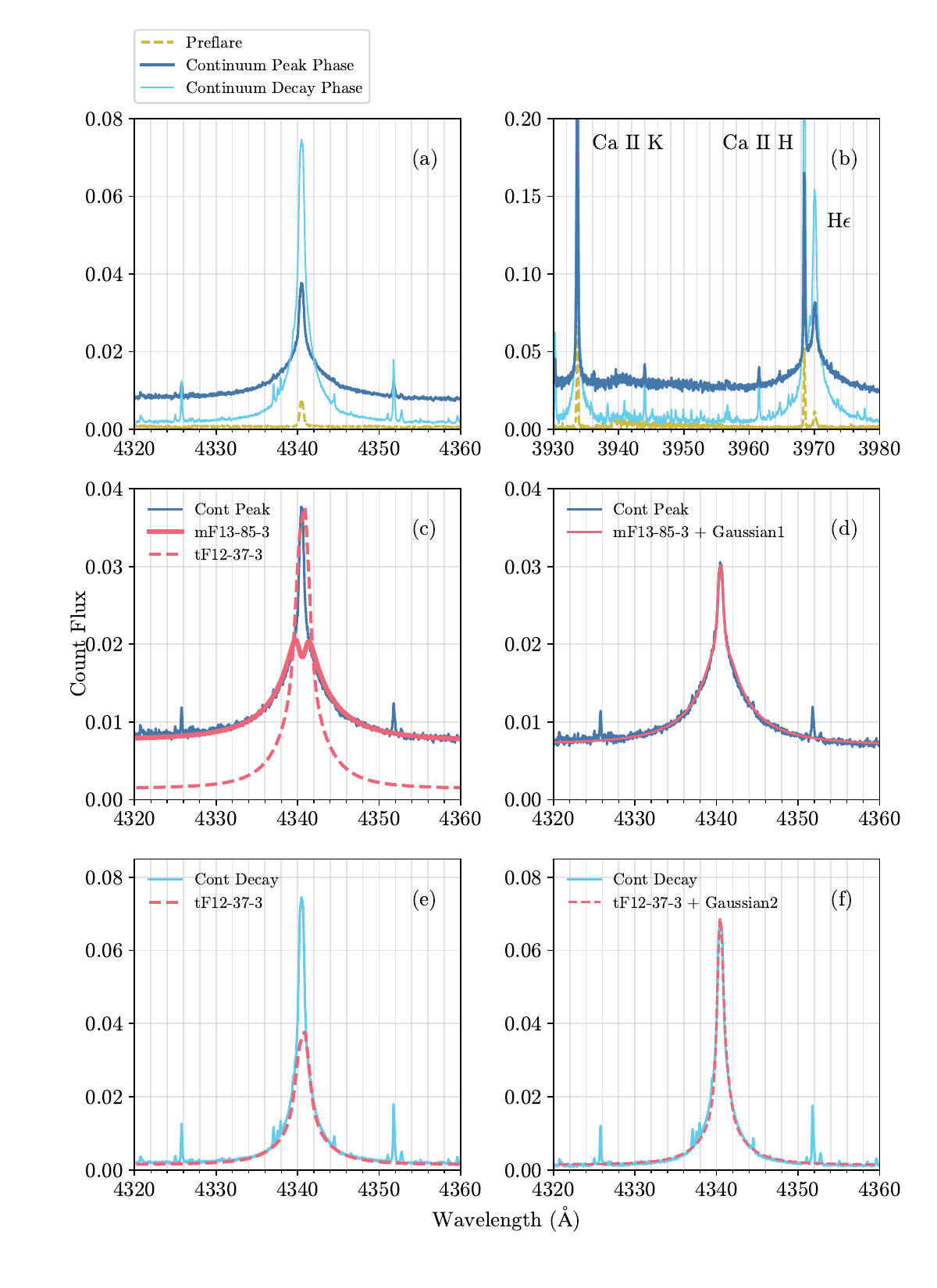}
		\caption{ Model spectra, $F_{\lambda}$,  of the H$\gamma$ emission lines during a flare on YZ CMi.   \textbf{(a)} Observations of the Balmer H$\gamma$ line in three consecutive spectra.  \textbf{(b)} Observations of Ca II K, Ca II H, and Balmer H$\epsilon$.  \textbf{(c)} The \texttt{mF13-85-3} model spectrum  \citep{Kowalski2022Frontiers} well-reproduces the H$\gamma$ wing shape and the relative flux of the nearby continuum radiation in the continuum peak phase observation (dark blue line).  \textbf{(d)}  An additional narrow Gaussian well-reproduces the entire H$\gamma$ line profile.  \textbf{(e)}   The \texttt{tF12-37-3} model spectrum \citep{Namekata2020} well-reproduces the H$\gamma$ wing shape and the relative flux of the nearby continuum radiation in the continuum decay phase observations (light blue line).  \textbf{(f)}  An additional narrow Gaussian well-reproduces the entire H$\gamma$ line profile.   In panels (d) and (f), the pre-flare observation has been subtracted from the flare data.  The RHD model surface fluxes have been scaled to the count fluxes (in units of counts s$^{-1}$) of the nearby continuum levels in the spectral observations.  The model spectra with $n\lambda=327$ (Section \ref{sec:spectracalc}) are used in this figure.
\label{fig:Vida}}
\end{center}
\end{figure}

Spectra around Balmer H$\gamma$ and H$\epsilon$ are displayed in Figure \ref{fig:Vida}(a) and Figure \ref{fig:Vida}(b), respectively.  The sequential exposures include a pre-flare spectrum, a spectrum with a powerful response throughout the optical continuum and highly broadened hydrogen Balmer lines, and a spectrum with larger line-to-continuum ratios  and less broadened hydrogen lines.  We refer to the two flare spectra as the ``continuum peak'' and the ``continuum decay'' spectra.  From the continuum peak to the continuum decay spectra, the value of  $F_{\rm{H}\gamma}^{\prime}$ /C4170$^{\prime}$ increases from  $\approx 10$  to $\approx 100$, and the effective width (Eq. \ref{eq:effective_width}) of the H$\gamma$ line decreases from $\Delta \lambda_{\rm{eff}} = 4.5$ \AA\ to $2.5$ \AA.    Although broadband photometry data are not available to provide context for the spectra, it is conceivable that the long exposure times integrated over flare phases similar to the echelle observations of the giant flare that was analyzed in \citet{Fuhrmeister2008}.  It is also possible that the Balmer line flux peaks after the continuum, an effect that occurs in some events \citep{HP91, Garcia2002, Namekata2020}.

In Figure \ref{fig:Vida}(b), the H$\epsilon$ line broadens far more than the Ca II H and K lines. Such differences are well understood to be the result of enhanced charged-particle pressure (Stark) broadening of the hydrogen lines in high-density flare chromospheres \citep{HP91, Allred2006, Paulson2006, Kowalski2022}.  We compare to the predictions of the \texttt{mF13-85-3} model around the H$\gamma$ line and the nearby continuum flux (Figure \ref{fig:Vida}(c)). The model spectrum, $F_{\lambda}$, at $n\lambda=327$ wavelength points (Section \ref{sec:spectracalc}) is multiplied by a scale factor so that it matches the approximate continuum count fluxes around $\lambda \approx 4320$ and $\lambda \approx 4360$ \AA. The comparison demonstrates that  the shape of the wings is reproduced, which is remarkable because the model spectrum is a temporal average that consists of the evolution of the model atmospheric densities, temperatures, and heating rates (Section \ref{sec:quantities}).  The Balmer H$\beta$ and H$\delta$ model predictions are also within the constraints of the observed wings of these lines (not shown).  The H$\gamma$ profile from the \texttt{tF12-37-3} model in comparison demonstrates that lower energy beam heating does not reproduce the relative fluxes in the wings and nearby continuum. This model is a much better match to the continuum decay phase spectrum (Figure \ref{fig:Vida}(e)).

The goal of modeling flares with high-flux electron beams and large values of $E_c$ is to reproduce the optical continuum properties in M dwarf flares.  Then, we assess the extent to which such models agree with emission line spectra.   Clearly, the electron beam heating models in Figure \ref{fig:Vida} do not account for the narrow, core emission in H$\gamma$.  The discrepancy is especially evident for the comparison of the continuum peak phase observation to the mF13-85-3 model (Figure \ref{fig:Vida}(c)), which has central dip at line center. The reason for the central dip will be clarified in Section \ref{sec:formation}.  A least-squares fit of a Gaussian to the extra amount of line core emission in the observations gives best-fit FWHM values of $0.7-0.8$ \AA , and the decay phase spectrum requires a Gaussian with a factor of $\approx 2.5$ greater flux at Earth.  The total models, which each consist  of a primary beam model spectrum and a Gaussian, are shown in Figures \ref{fig:Vida}(d,f).  In \citet{Kowalski2022Frontiers}, a similar discrepancy in the Balmer lines was rectified by superposing a lower beam flux model with one of the \texttt{mF13-85-3}, \texttt{mF13-150-3}, or \texttt{mF13-500-3} models.  The interpretation is that lower and higher-energy electron beams impact different areas \citep[see also][]{Cram1982, Kowalski2010}, which is in line with detailed analyses of optical  \citep{Neidig1993} and near-ultraviolet \citep{Kowalski2017Broadening} solar flare kernels at high spatial resolution. However, only a few of our models (\texttt{mF10-150-3} and \texttt{mF10-17-3}) produce H$\gamma$ line profiles with FWHM values that are as small as the Gaussian fits suggest. To account for the narrow line core flux at Earth, the area of the  region emitting the \texttt{mF10-17-3} spectrum would have to be a factor of $6.5$ larger than the source area with the higher beam flux heating (\texttt{mF13-85-3}) that accounts for the continuum and the far wing radiation in the impulsive phase.

\subsubsection{Line Formation Properties} \label{sec:formation}
The \texttt{mF13-85-3} and \texttt{tF12-37-3} models satisfactorily explain the H$\gamma$ broadening at $|\lambda - \lambda_0| \gtrsim 0.5$ \AA\ in the continuum peak and continuum decay phase flare spectra, respectively.  In this section, we compare the formation of the H$\gamma$ lines in these models.  We analyze the \texttt{tF12-37-3} model, instead of the very similar \texttt{mF12-37-3} model, because 
 \citet{Namekata2020} used this simulation to understand the formation of the broad H$\alpha$ lines in a superflare on AD Leo.

\begin{figure*}
\gridline{\fig{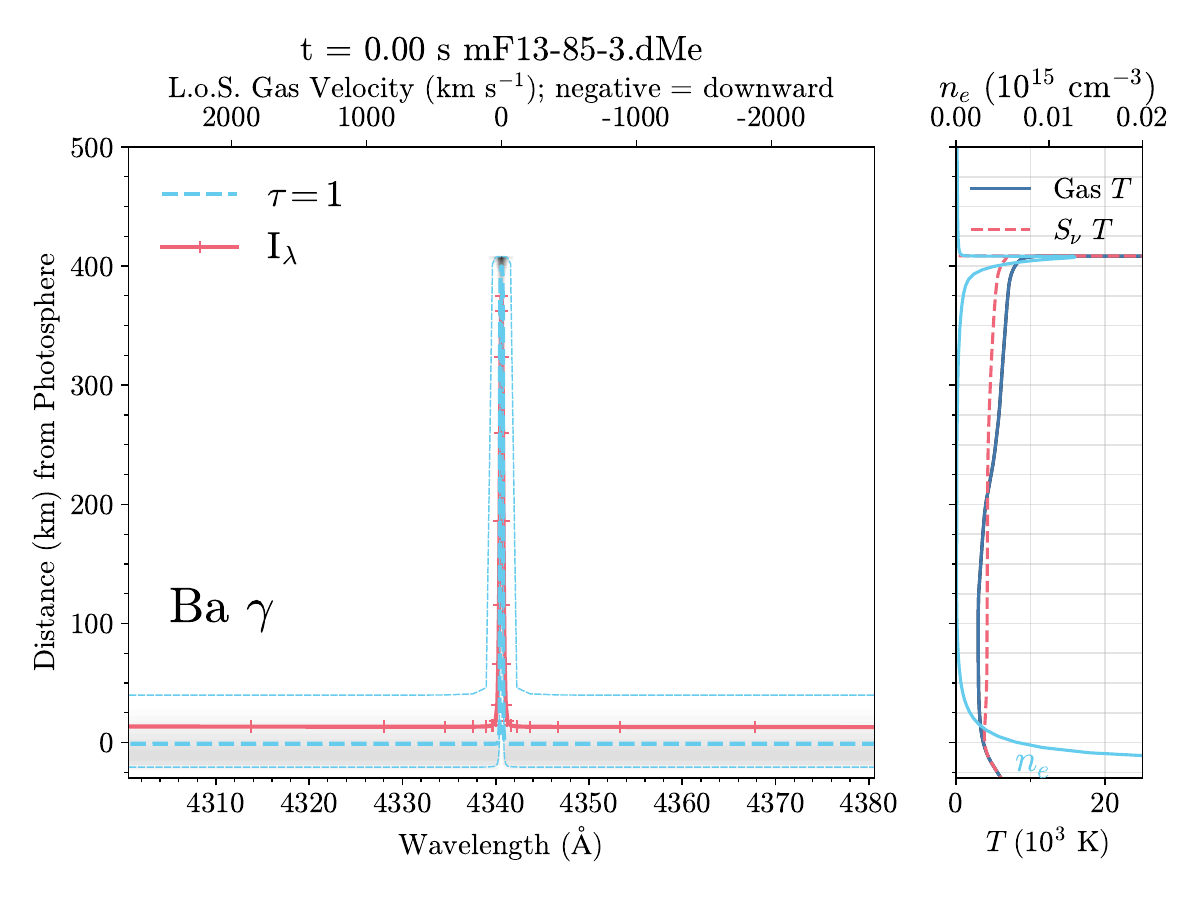}{0.45\textwidth}{(a)}
          \fig{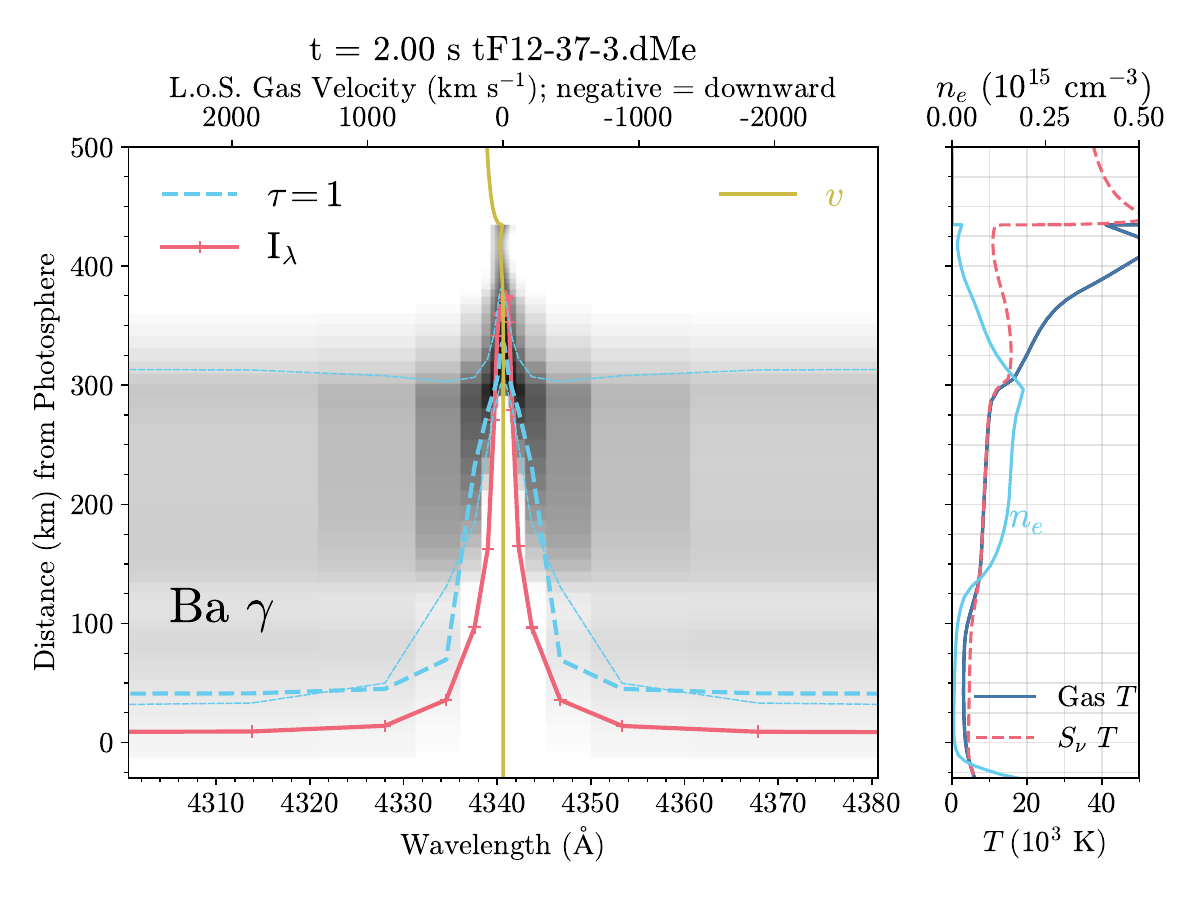}{0.45\textwidth}{(b)}
          }
\gridline{\fig{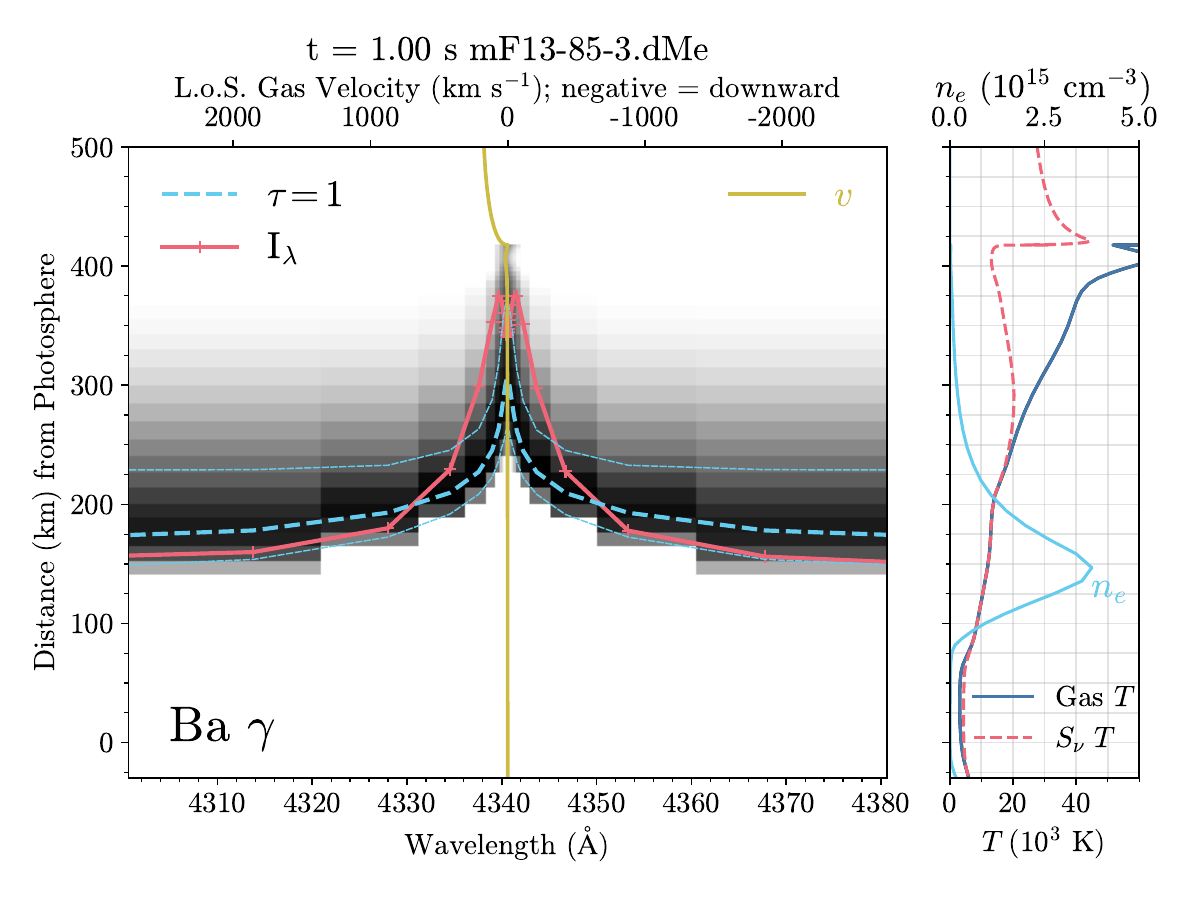}{0.45\textwidth}{(c)}
          \fig{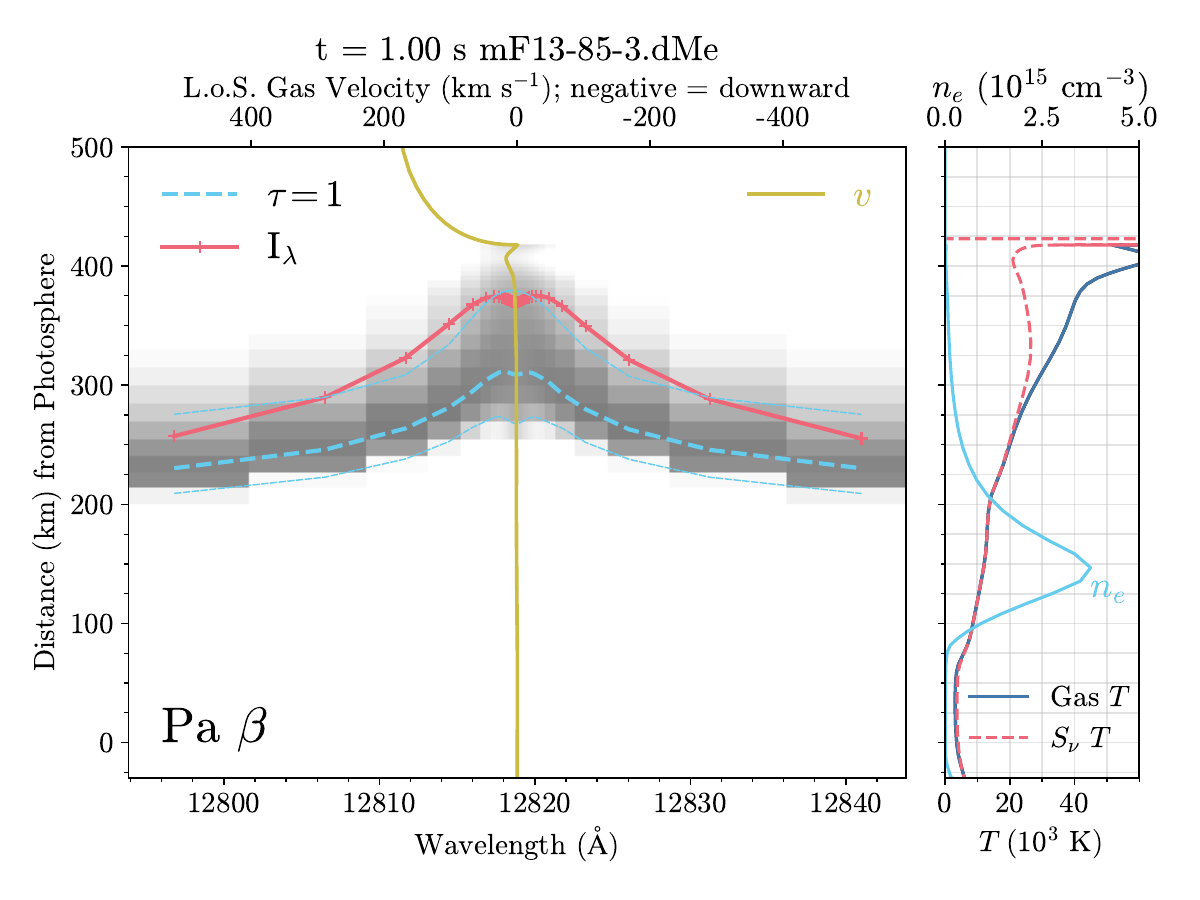}{0.45\textwidth}{(d)}
          }          
\caption{ Contribution functions to the emergent $\mu=0.95$ intensity for the hydrogen Balmer (Ba) $\gamma$ line at $t=0$~s (a), at  $t=2$~s during the \texttt{tF12-37-3} model (b), and at $t=1$~s during the \texttt{mF13-85-3} model (c).  Panel (d) shows the contribution function over the hydrogen Paschen (Pa) $\beta$ line at $t=1$~s during the  \texttt{mF13-85-3} model.
The gray scale is logarithmic from $10^{-2}$ (white) to $10^{1.5}$ (black) erg cm$^{-2}$ s$^{-1}$ sr$^{-1}$ \AA$^{-1}$ cm$^{-1}$ in all panels.  The thin dashed (blue) lines 
indicate 5\% (upper) and 95\% (lower) of the cumulative of the contribution function, and the top axes indicate the line-of-sight (L.o.S.) velocities that correspond to the wavelength detuning and to the gas velocity (yellow line).   Margin plots show the source-function equivalent temperatures ($S_{\nu}$ $T$) and the gas temperatures (bottom axes) compared to the ambient (thermal) electron densities (top axes).  Note the different scales among the electron density axes. As noted in \citet{Kowalski2022Frontiers}, the \texttt{main} \texttt{mF13} models produce non-negligible chromospheric upflows, but these are not apparent on the velocity scale shown here.
\label{fig:ContribFunct}}
\end{figure*}

Figure \ref{fig:ContribFunct} shows contribution functions to the emergent radiative intensity across the hydrogen Balmer $\gamma$ line.
  The preflare formation of the model Balmer $\gamma$ intensity (Figure \ref{fig:ContribFunct}(a)) is displayed for the same wavelength and color scales as for the flare simulations (Figure \ref{fig:ContribFunct}(b-c)) in order to emphasize the amounts of the pressure broadening increase in the high-flux beam models.  The thin dashed (blue) lines indicate the heights corresponding to 5\% and 95\% of the emergent intensity, and the thick dashed (blue) lines show the $\tau(\lambda)=1$ heights.  The margin plots in each panel show the electron density, gas temperature, and the source-function equivalent temperature.  Comparisons of these two measures of temperature indicate where the line and nearby continuum formation is close to local thermodynamic equilibrium (LTE) conditions.  In both models, the line core formation deviates from LTE.  In the \texttt{mF13-85-3} model, the line core has a central ``absorption'' because the source function equivalent temperature decreases over a $\Delta z \approx 100$ km height range following its departure from the gas temperature curve.  The line core is formed over much lower electron densities than the wings \citep{Kowalski2017Broadening, Namekata2020}.
  In the wings and nearby continuum, the \texttt{mF13-85-3} model becomes optically thick at electron densities of $n_e \approx 1-2.5 \times 10^{15}$ cm$^{-3}$ at heights well above the heights that correspond to $\tau(\lambda)=1$ in the preflare and in the \texttt{tF12-37-3} models.  The formation of the Balmer $\gamma$ line wings in the \texttt{mF13-85-3} model is very close to LTE.
  
\begin{figure}
\begin{center}
\includegraphics[scale=.75]{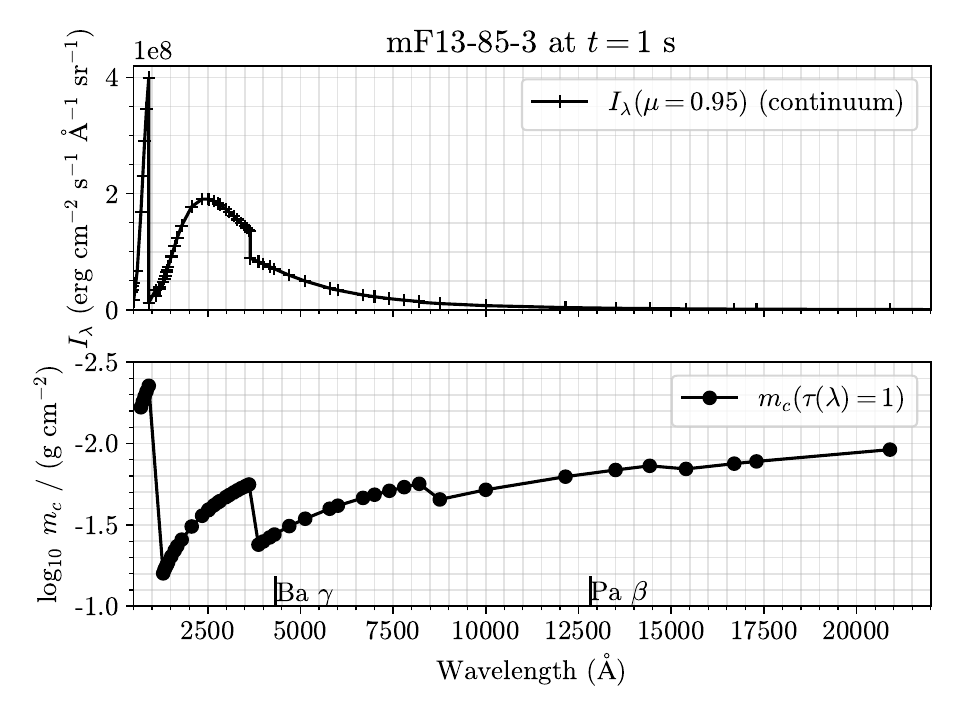}
\caption{ (top) Emergent continuum intensity spectrum ($\mu=0.95$) for the \texttt{mF13-85-3} model at $t=1$~s.  (bottom)  Column mass corresponding to $\tau(\lambda)=1$ for a representative subset of 53 continuum wavelengths from the top panel between $\lambda=700$ and $\lambda = 20900$ \AA.  In the bottom panel, the central wavelengths for hydrogen Balmer $\gamma$ and Paschen $\beta$ are indicated (see text).
\label{fig:Tau1Cont}}
\end{center}
\end{figure}

  The formation of the hydrogen Paschen $\beta$ ($n=3$ to $n=5$) transition is shown in Figure \ref{fig:ContribFunct}(d) for the \texttt{mF13-85-3} model at $t=1$~s. This emission line shares the same upper level as Balmer $\gamma$ and thus should experience approximately the same amount of Stark broadening.   This transition is also more optically thin than the Balmer $\gamma$ transition, which ostensibly allows one to probe deeper into the atmosphere  \citep{Kowalski2022}, where there are larger electron densities in the \texttt{mF13-85-3} model. Figure \ref{fig:Tau1Cont}(top) shows the emergent continuum intensity at the wavelengths calculated in detail.  The contribution functions and optical depths at continuum wavelengths are obtained from the model output following the method of \citet{Kowalski2017Mar29}. Figure \ref{fig:Tau1Cont}(bottom) indicates the column mass at which $\tau=1$ occurs at selected wavelengths in the top panel. The continuum wavelengths ($\lambda \approx 12150$ \AA) near Paschen $\beta$ are  more optically thick at a given height in the atmosphere than the blue continuum wavelengths ($\lambda \approx 4300$ \AA) around Balmer $\gamma$.  This pushes the formation of the Paschen $\beta$ line to smaller column masses in the  atmosphere with higher temperatures and lower electron densities (at this particular time in this model).  The source function equivalent temperature for the Paschen $\beta$ line is closer to the gas temperature, indicating that it is formed closer to LTE than Balmer $\gamma$, as expected.

\begin{figure}
\begin{center}
\includegraphics[scale=.75]{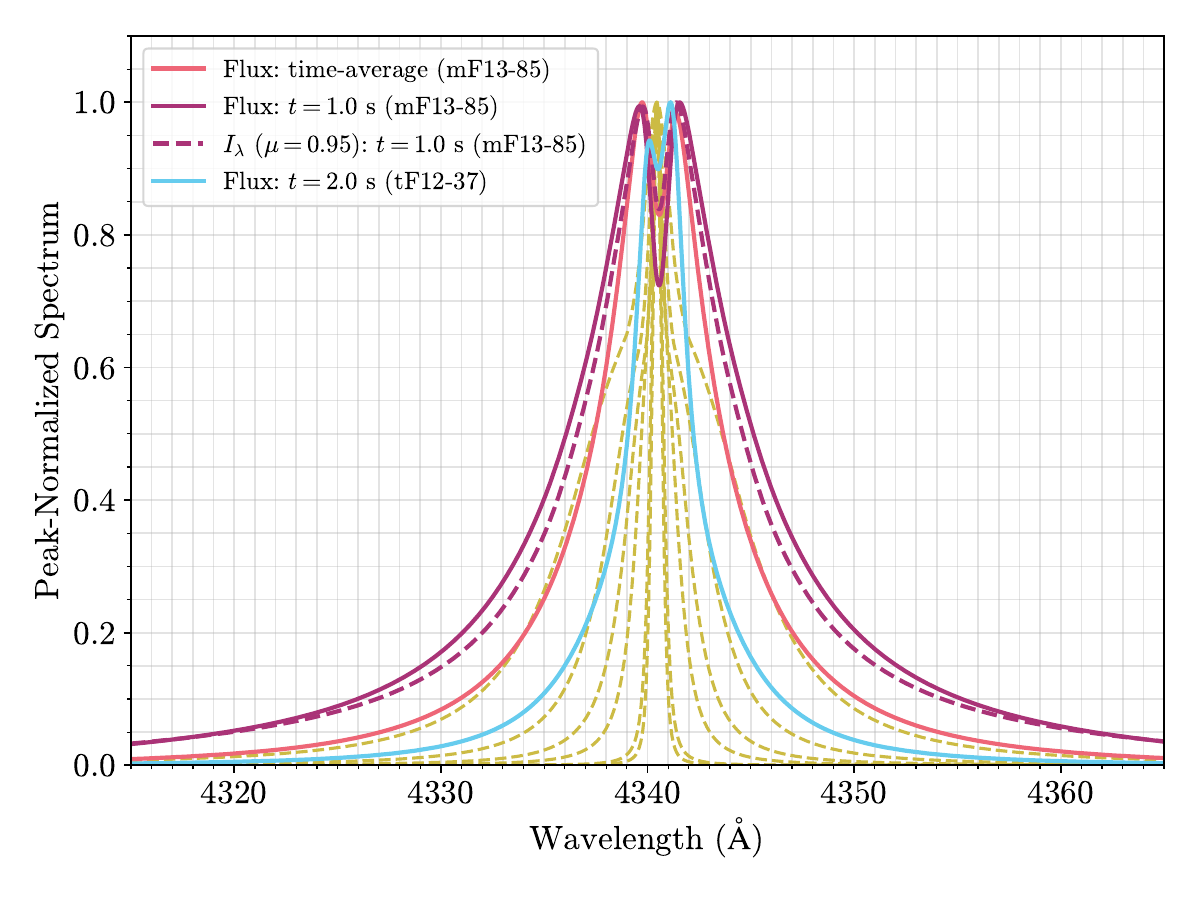}
\caption{ Model spectra around the hydrogen Balmer $\gamma$ line in the two RHD models (\texttt{mF13-85-3} and \texttt{tF12-37-3}) from Figure \ref{fig:Vida}. The $y$-axis is the integrand of Eq. \ref{eq:effective_width}:  the nearby continuum has been subtracted, and the resultant spectra are normalized to the maximum over the line.   Doppler-broadened TB09$+$HM88 line profile functions, which are equivalent to optically thin static slab calculations when peak-normalized,  are shown as progressively broader dashed yellow spectra for electron densities of $n_e = 5 \times 10^{13}, 10^{14}, 5\times10^{14}, 10^{15}, 2\times10^{15},$ and $5\times10^{15}$ cm$^{-3}$ at $T=10^4$ K.
\label{fig:HgammaSpec}}
\end{center}
\end{figure}

To accurately interpret electron densities from the pressure broadening of spectral lines in flares it is crucial to account for the amount of broadening due to optical depths.   Optical depth effects in flare models can greatly enhance the broadening beyond the expectation from the largest electron density in the flare atmosphere \citep[e.g.,][]{JohnsKrull1997, Kowalski2022}.
How well does the amount of broadening in hydrogen Balmer lines relate to the actual electron densities achieved (\emph{cf.} the margin plot in Figure \ref{fig:ContribFunct}(c)) in the \texttt{mF13-85-3} flare atmosphere?  Figure \ref{fig:HgammaSpec} compares continuum-subtracted, peak-normalized spectra of Balmer $\gamma$.  Here, we use the spectra calculated at $n\lambda=327$ wavelength points (Section \ref{sec:spectracalc}). The \texttt{mF13-85-3} model is compared to several spectra that are calculated from optically thin slabs with a uniform electron density.  The two broadest slab model spectra are calculated with $n_e=2 \times 10^{15}$ cm$^{-3}$ and $n_e=5\times 10^{15}$ cm$^{-3}$, respectively.  The emergent intensity spectrum at $t=1$~s from the \texttt{mF13-85-3} model is much broader than these, even though the electron densities in the atmosphere are  far below $n_e=5\times 10^{15}$ cm$^{-3}$ (Figure \ref{fig:ContribFunct}(c)).
This discrepancy is due to the large optical depths ($\tau \approx 1$) over the heights where the line is formed.    The flux spectrum from the \texttt{tF12-37-3} model at maximum beam injection ($t=2$~s) is as broad as the $n_e=2\times10^{15}$ cm$^{-3}$ slab calculation, but the actual electron densities where the line is formed are an order of magnitude smaller\footnote{All of the discrepancies that were discussed remain if the \texttt{RADYN} model spectra are scaled to 0.75 instead of 1.0.} (Figure \ref{fig:ContribFunct}(b)).

It is also valuable to compare the time-averaged flux profile ($F_{\lambda}$) that fits the Balmer wings in the YZ CMi flare (Figure \ref{fig:Vida}) to the intensity and flux spectra ($F_{\lambda}(t=1~s)$) at the time of maximum beam heating.
The time-averaged flux spectrum from the \texttt{mF13-85-3} model is narrower than the $t=1$~s flux spectrum because the chromosphere experiences the deepest heating and largest ionization around the time of the maximum injected beam flux at $t=1$~s (Figure \ref{fig:Figure_BeamEvol}).  The $t=1$~s flux and intensity spectra exhibit minor differences in the wing shapes and larger differences in the core\footnote{This is expected due  to the plane-parallel calculation of the radiative transfer.  At smaller $\mu$ values, the optical depth accrues to larger values at a given height;  the formation of the core then occurs at smaller values of the source function.}.

\section{Discussion} \label{sec:discussion}

We have calculated a grid of 80 RHD stellar flare models with the \texttt{RADYN} code.    The \texttt{RADYN} models extend to large electron beam fluxes in order to reproduce the empirical optical and near-ultraviolet continuum properties in M-dwarf flares.  Small electron beam fluxes are included too;  these do not reproduce any property to which we compare unless the  low-energy cutoff is very large.  For large low-energy cutoffs and large beam fluxes, scaling to the optical (or NUV continuum, if available) reproduces the $T \gtrsim 10^4$ K blackbody color temperatures and small Balmer jumps in the impulsive phases of several well-studied M dwarf flares.  We have shown that if a continuum prediction from one such model is scaled to the observed optical continuum flux in a large dMe flare, the wing shape of the highly broadened H$\gamma$ line is remarkably well-reproduced.  

 The large low-energy cutoffs in our RHD modeling approach imply that large heating rates are needed over low chromospheric heights in stellar flares.  One possible source of deep heating that is not often considered arises from MeV proton beams \citep{Rieger1996, Emslie1998, Ondrej2, Allred2020, Sadykov2024}.  Another possibility is that time-dependent transport effects \citep{Kontar2012} generate an enhancement of $E \gtrsim 100$ keV electrons in stellar flares. The enhancement produces a similar local heating maximum at $m_c \approx 0.01$ g cm$^{-2}$ as in the  $E_c = 85$ keV  models with $\delta = 3$, and the temperatures around this column mass exceed $T = 10^4$ K if the energy flux is large enough \citep{Kowalski2023}.  We note that in some solar flares, there is growing evidence that deeper heating is needed beyond that from single-power law electron beams that are inferred from the standard interpretation of nonthermal hard X-rays \citep{Martinez2012, Warren2014, Ondrej1, Kowalski2019IRIS}.

The grid of electron beam models that are listed in Table \ref{table:calculations}, which we refer to as the v1.0 models, simulate flare heating ($Q_{\rm{flare}}$) from binary, beam-plasma Coulomb collisions only.  Collective forces on the beam particles have not yet been included in our M dwarf flare models.  The effects of return current energy loss and magnetic mirroring  introduce a large number of additional free parameters \citep[e.g., turbulent scattering effects and  the location of magnetic field convergence;][]{Murphy2007, Kontar2014}.  In the state-of-the-art computational framework \citep{Allred2020}, the return current and magnetic forces are limited to the steady-state solution of the distribution function.  This assumption may be valid, but more work is needed to determine how time-dependent transport effects \citep[e.g.,][]{Kontar2012} alter the steady-state distribution.  Additionally, nonthermal runaways may be generated in the background plasma due to large return current electric fields \citep{Holman1985, Alaoui2021}.  Nonetheless, it is expected that the summative effects of the return current electric field would prevent much of the beam energy in our models from reaching the low chromosphere, thus precluding any electron beam model from achieving the deep heating that explains the stellar observations.

The v1.0 model grid has a large number of valuable applications.  These models have already been leveraged in a variety of ways:  studies of the  merging of the Balmer series \citep{Kowalski2017Broadening} for NUV exoplanet habitability assessments \citep{Kowalski2022Frontiers}, the energy partition in NUV and optical photometry \citep{Brasseur2023}, infrared flaring in JWST spectra and the response of the He I $\lambda$10830 line \citep{Howard2023}, solar-stellar comparisons \citep{Monson2024}, and the relations among gas, color, and radiation temperatures in flare models \citep{Kowalski2023}.   Many further applications of the grid models are possible.  For example, the Balmer line profiles in the deep heating models can be comprehensively tested against a recently published large sample of M dwarf flares with high time-resolution and broadband photometry \citep{Notsu2024}.  Other uses include the following: FUV and Balmer continuum slopes can be compared to observations \citep{Loyd2018Hazmat, Froning2019, Kowalski2019HST, Feinstein2022, Chavali2022}, optically thin coronal and transition region emission lines can be calculated \citep{Allred2006} from the early stages of chromospheric evaporation, the relative light-curve timescales between the FUV continuum and the $U$-band  \citep{Hawley2003} can be investigated,  more gradual-type white-light events with larger Balmer jumps (Figure \ref{fig:ColorColor}) can be modeled in their impulsive phases \citep{Kowalski2019HST}, and the broadening of the Paschen lines \citep{Fuhrmeister2008, Schmidt2012, Kanodia2022, Howard2023} can be modeled in detail.  It is also possible to readily synthesize optically thick UV resonance lines of C II, Mg II, and possibly Si IV \citep{Kerr2019Si} with a complementary radiative transfer code  \citep[e.g.,][]{Uitenbroek2001, Kerr2019A, Zhu2019}.

 The expansive range of input heating rates in the grid facilitates  semi-empirical modeling of stellar flares with two beam components \citep[e.g., a linear superposition of model spectra through maximum likelihood parameter estimation;][]{Kowalski2022Frontiers}. 
We applied the concept to a large dMe flare with archival echelle observations of the hydrogen Balmer $\gamma$ line (Section \ref{sec:detailed}).  The impulsive-phase and gradual-phase blue-optical continuum levels together with the H$\gamma$ wing shapes are well-reproduced by high-energy beam models from the grid.  In solar flare multithread models that use thermal conduction pulses \citep{Warren2006}, the post-impulsive phase is explained using a superposition of equal-length loops, and the energy flux deposited per loop decreases as the flare progresses. Recently, \citealt{Rempel2023} showed the evolution of thermally conducted energy fluxes into the footpoints in one of their 3D solar flare simulations.  Although a comparison to 1D multithread models \citep[e.g.,][]{Warren2006} was not discussed, it appears that a similar progression occurs with higher average fluxes into the footpoints during the earlier evolution. The differences in the energy fluxes of high-energy electron beams between the \texttt{mF13-85-3} model and the \texttt{tF12-37-3} represents an analogous change from peak to decay in the YZ CMi flare studied here.  These two primary beam models are remarkably different, which is justified in accounting for the \emph{striking} differences in the spectral properties (line-to-continuum ratios, line wing shapes, effective widths) in the observations.  The modeling of the continuum and wing in each of the observations further suggests an evolution in the average energy per beam particle from $\approx 170$ keV to $\approx 75$ keV.  As the flare reconnection  progresses higher into the corona (which occurs in loss-of-equilibrium models, such as in Figure \ref{fig:geometry}), the flux and average energy per particle may sensibly decrease with the decreasing strength of the magnetic field, which must overcome the increasing length of the current sheet \citep{Lin2000, Reeves2005}. All of this supports the idea that the evolution of spatially-integrated stellar flare light curves is not solely driven by changing ribbon areas or cooling times of monolithic loops.

   Nonthermal electrons leave signatures in the microwaves well into the gradual decay phase of the white-light in stellar flares \citep{Osten2005} and after the peak of the hard X-rays in solar flares \citep{Krucker2020}.  So, one could imagine that there is a relationship between  gyrosynchrotron-emitting electrons and chromospheric heating after the impulsive phase.  Multi-wavelength analyses of the empirical optical-radio connection in stellar flares \citep{Osten2016, Tristan2023} would be helpful because the high-flux model (\texttt{tF12-37-3}) component in the decay phase of the continuum radiation implies that accelerated electrons are important in atmospheric heating even at late times \citep[see also the \texttt{RADYN} multithread modeling of the decay phase of a megaflare in][]{Kowalski2017Broadening}. However, previous spectral observations of the decay phase in stellar flares show that the Balmer jump ratios are even smaller than the prediction of the \texttt{tF12-37-3} model (\emph{cf.} Fig 21 of \citealt{Kowalski2013}).  
   
  We leave the origin of the narrower and fainter line-core model components in each of the impulsive and gradual phase spectra in Figure \ref{fig:Vida} to future work when other lines can be incorporated into a comprehensive statistical analysis with appropriate priors on the Balmer jump ratios (which are not available in the echelle spectra from NARVAL).  The origin of the late-phase  peak in Ca II K flux in stellar flares \citep{HP91,Garcia2002,Kowalski2013} is yet unexplained \citep{Kowalski2022Frontiers}, and detailed comparisons to the non-hydrogen emission lines in the observations (e.g., Figure \ref{fig:Vida}(b)) would thus be valuable. However, there may be unavoidable limitations in the 1D grid of models that we present here. In the \citet{Reale1997} paradigm for modeling the soft X-ray emission of stellar flares, the inferred loop lengths are much larger than 10 Mm \citep{Gudel2004, Reale2004, Osten2010, Osten2016}. In our grid, the short durations ($\Delta t \approx 10$~s) over which the coronal plasma in a small loop are followed may be partly responsible for the discrepancies around the emission line cores.
  For example, coronal irradiation from a large volume of late-phase flare loops might be able to sustain Ca II at high levels into the gradual decay phase of the optical and NUV continuum radiation \citep{HF92}.  Alternatively, we speculate that a temperature increase due to betatron heating / acceleration \citep{Veronig2006, Birn2017} during a gradual phase of shrinkage of each loop might contribute to heating that is required to raise the  source function around formation layers of the Balmer and Ca II line cores in high-energy, electron-beam models.

  The models in our grid are intended to reproduce the bulk of the optical and near-ultraviolet radiation in the impulsive-phase of a variety of dMe flare energies, peak amplitudes, and durations (Figure \ref{fig:ColorColor});  they are particularly well-suited for the rise and peak phases of flares that exhibit evidence for hot blackbody-like optical continuum radiation and relatively small Balmer jump ratios \citep[e.g.,][]{Kowalski2013, Kowalski2023}. In the early hard X-ray and white-light impulsive phase of solar flares, brightenings  elongate the size of ribbons, which tend to closely straddle the magnetic polarity inversion lines \citep[e.g.,][]{Qiu2010, Qiu2017, Kazachenko2022, Tamburri2024}.
  The projected areas of loop arcades can grow substantially over time \citep[e.g.,][]{Aschwanden2012Logistic} as the ribbons  spread apart. Thus,  chromospheric ribbons in the early phase are connected by the shortest loops within a flare (in the absence of large shear initially). The length chosen for a model loop in the early phase of a flare does not have a large effect on the Coulomb energy losses from representative beam electrons in transit to the chromosphere:  for an electron density of $n_e=5\times 10^{10}$ cm$^{-3}$, the stopping distance \citep{Emslie1978} for a beam electron with initial kinetic energy $E_0 =75$ keV is 170 Mm, and the stopping distance for a $E_0 = 170$ keV beam electron is 760 Mm.  The temporal averaging of radiative flux spectra in models with short beam-injection timescales ($\Delta t \sim 2$~s)  approximates the superposition of many sequentially heated chromospheric kernels in stellar flares, which presumably also have an early phase of ribbon elongation and relatively small loop lengths connecting conjugate footpoints. The increasing areas that are inferred through optical blackbody fitting in the rise phase of stellar flares generally supports this analogy \citep[e.g.,][]{Kowalski2013}.   
 
  We can further justify the short timescales of beam heating by comparing to a recent high-resolution analysis of a solar flare.   \citet{Graham2020} calculated that significant amounts of newly brightened chromospheric areas in a $\approx 10^{32}$ erg solar flare occur within an observational cadence of 19~s, which is a factor of $\approx 10-20$ shorter duration than the total hard X-ray impulsive phase duration.  Crudely extending this relation to the $U$-band impulsive phase timescales (which we take as the $t_{1/2}$ values in  Table 6 of \citet{Kowalski2013} and Table 6 of \citealt{Kowalski2016}) of the stellar flares in Figure \ref{fig:ColorColor} suggests that it is reasonable to assume that beam heating lasts for only several seconds  in each impulsive-phase loop.  Better assessments that leverage unique capabilities of the ROSA instrument \citep{Rosa} are forthcoming (Kowalski et al 2024, in preparation).

\section{Conclusions} \label{sec:conclusions}
There has been an explosion of stellar flare observations in the last few decades, but modeling has generally been limited to static, semi-empirical atmospheric calculations or to slabs.  When confronted with white-light constraints, time-dependent models (including 2D and 3D magnetohydrodynamic simulations) of X-ray flares do not make predictions at all.  The \texttt{RADYN} M-dwarf flare model grid is a modern extension of the static, semi-empirical atmospheric flare models of \citet{Cram1982}, which (to our knowledge) were the first models to produce blackbody temperatures of $T \approx 14,000$ K and broad hydrogen Balmer emission lines.   The \texttt{RADYN} models include vertical atmospheric heterogeneity and optical depths that are self-consistent with the equations of RHD in response to Coulomb heating from a power-law distribution of electrons.  We analyze observations using superpositions of time-dependent models, which emulate lateral heterogeneity over the spatially unresolved flare regions on other stars. The models reproduce the full range of Balmer jump properties in the observations of the impulsive phase, and we discuss how they are also useful for understanding the role of nonthermal particle heating in the gradual decay phases of light curves of the optical and NUV continuum radiation. Most notably, the hydrogen Balmer $\gamma$ wing shape that results from very high rates of deep atmospheric heating agrees with  symmetric broadening in  echelle flare observations (Figure \ref{fig:Vida}(c)), thereby providing a plausible physical interpretation 
of impulsive-phase, Balmer line ``broad components'' that are often phenomenologically modeled with a wide Gaussian  \citep[e.g.,][]{Doyle1988, Fuhrmeister2008}.  In the impulsive phase models, the Balmer wings are formed over large atmospheric electron densities and optical depths, from which the observed blue continuum properties also follow.  We conclude that models with deep atmospheric heating provide a robust paradigm that improves upon the ubiquitous $T \approx 9000$ K blackbody interpretation of stellar flare white-light continuum radiation.

\appendix
\section{Model Grid Access} \label{sec:appendix}
In this appendix, we describe how to access the model grid output.  FITS binary tables contain the detailed spectra and atmospheric variables at $\Delta t = 0.2$~s (in most cases).  Average spectral quantities, such as those in Table \ref{table:calculations}, are contained in \verb|modelvals.tave.fits|.   We distribute a supplementary Jupyter notebook that demonstrates how to load the outputs of the models.  The Python routines and this notebook can be obtained through PyPI by issuing the following command:  \verb|pip install radyn_xtools|.  We recommend consulting the documentation on the PyPI webpage for \verb|radyn_xtools| about setting up a standalone conda environment. The user is also advised to consult the \texttt{RADYN} documentation and IDL routines that are linked to the F-CHROMA solar flare model grid paper \citep{Carlsson2023}.  The original Common Data Format files are large in size, but they are available upon request to the corresponding author.

 The {\tt .fits} files along with the documentation that describes the installation of {\tt radyn\_xtools} and the contents of the files are available at the Zenodo repository, \url{https://doi.org/10.5281/zenodo.10929515}.

\acknowledgments
We thank an anonymous referee for helpful comments and critiques.
AFK thanks Dr. Kosuke Namekata for providing models used in his paper, Dr. Rachel Osten for discussions about grid applications and nonthermal particles in stellar flares, Dr. Jim Drake and Dr. Eduard Kontar for conversations about beam instabilities and low-energy cutoffs, Dr. Pier Emmanuel-Tremblay for providing the hydrogen broadening calculations, Dr. Thomas Gomez for helpful conversations about line shape theory, and Dr. Dana Longcope for helpful discussions about magnetic field retraction.  AFK thanks Dr. Graham Kerr and Dr. Chris Osborne for recommending Plotly. CHIANTI is a collaborative project involving the University of Cambridge (UK), the NASA Goddard Space Flight Center (USA), the George Mason University (GMU, USA) and the University of Michigan (USA).  AFK acknowledges funding support from the award NASA: 80NSSC21K0632.

\bibliography{main_v7}{}
\bibliographystyle{aasjournal}

\end{document}